\documentclass{SCIS2021}

\usepackage[english]{babel}
\usepackage{amsthm}
\usepackage{multirow} 
\usepackage{array}
\usepackage{wasysym}
\usepackage[colorlinks,linkcolor=black]{hyperref}
\newtheoremstyle{exampstyle}
{0.0em} 
{0.0em} 
{} 
{1em} 
{\bfseries} 
{.} 
{1em} 
{} 
\usepackage{balance}
\theoremstyle{exampstyle}

\usepackage{CJKutf8}
\usepackage{longtable}
\usepackage{makecell}
\usepackage{graphicx}



\begin{document}
	\ArticleType{RESEARCH PAPER}
	\Year{2021}
	\Month{}
	\Vol{}
	\No{}
	\DOI{}
	\ArtNo{}
	\ReceiveDate{}
	\ReviseDate{}
	\AcceptDate{}
	\OnlineDate{}
	
	\title{Towards Semantic-based Agent Communication Networks: Vision, Technologies, and Challenges}{Towards Semantic-based Agent Communication Networks: Vision, Technologies, and Challenges}
	
\author[1]{Ping ZHANG}{}
\author[1]{Rui MENG*}{{buptmengrui@bupt.edu.cn}}
\author[1]{Xiaodong XU*}{{xuxiaodong@bupt.edu.cn}}
\author[1]{Yaheng WANG}{}
\author[1]{Zixuan HUANG}{}
\author[1]{Yiming LIU}{}
\author[2]{\\Ruichen ZHANG}{}
\author[2]{Yinqiu LIU}{}
\author[3]{Haonan TONG}{}
\author[4]{Huishi SONG}{}
\author[5]{Gang WU}{}
\author[6,7,8]{Zhaoming LU}{}
\author[9]{\\Jiawen KANG}{}
\author[10]{Geng SUN}{}
\author[11]{Qinghe DU}{}
\author[12]{Zhaohui YANG}{}
\author[13]{Jingxuan ZHANG}{}
\author[14]{Han MENG}{}
\author[15]{\\Lexi XU}{}
\author[16]{Haitao ZHAO}{}
\author[17]{Zesong FEI}{}
\author[18,19,20]{Yiqing ZHOU}{}
\author[21]{Pei XIAO}{}
\author[22]{Meixia TAO}{}
\author[23]{\\Qinyu ZHANG}{}
\author[24]{Shuguang CUI}{}
\author[21]{Rahim TAFAZOLLI}{}
	\AuthorMark{ZHANG P}
	
		\AuthorCitation{ZHANG P, MENG R, XU X, et al}
	
	\contributions{Ping ZHANG and Rui MENG contributed equally to this work and should be considered co-first authors.}
	

\address[1]{State Key Laboratory of Networking and Switching Technology, BUPT, Beijing {\rm 100876}, China}
\address[2]{College of Computing and Data Science, Nanyang Technological University, Singapore {\rm 639798}, Singapore}
\address[3]{Aerospace Information Research Institute, Chinese Academy of Sciences, Beijing {\rm 100094}, China}
\address[4]{ZGC Institute of Ubiquitous-X Innovation and Applications, Beijing {\rm 100083}, China}
\address[5]{National Key Laboratory of Wireless Communications, UESTC, Chengdu {\rm 611731}, China}
\address[6]{Beijing Key Laboratory of Network System Architecture and Convergence, BUPT, Beijing {\rm 100876}, China}
\address[7]{Beijing Laboratory of Advanced Information Networks, BUPT, Beijing {\rm 100876}, China}
\address[8]{Xiong'an Aerospace Information Research Institute, Xiong'an {\rm 070001}, China}
\address[9]{School of Automation, Guangdong University of Technology, Guangzhou {\rm 510006}, China}
\address[10]{ College of Computer Science and Technology, Jilin University, Changchun {\rm 130012}, China}
\address[11]{ School of Information and Communications Engineering, Xi'an Jiaotong University, Xi'an {\rm 710049}, China}
\address[12]{ College of Information Science and Electronic Engineering, Zhejiang University, Hangzhou {\rm 310027}, China}
\address[13]{ National School of Elite Engineering, University of Science and Technology Beijing, Beijing {\rm 100083}, China}
\address[14]{ Institute of Network and IT Technology, China Mobile Research Institute, Beijing {\rm 100053}, China}
\address[15]{ Research Institute, China United Network Communications Corporation, Beijing {\rm 100048}, China}
\address[16]{ College of Electronic Science and Technology, National University of Defense Technology, Changsha {\rm 410073}, China}
\address[17]{ School of Information and Electronics, Beijing Institute of Technology, Beijing {\rm 100081}, China}
\address[18]{ State Key Lab of Processors, Institute of Computing Technology, Chinese Academy of Sciences, Beijing {\rm 100190}, China}
\address[19]{ Beijing Key Laboratory of Mobile Computing and Pervasive Device, Beijing {\rm 100190}, China}
\address[20]{ University of Chinese Academy of Sciences, Beijing {\rm 100049}, China}
\address[21]{ 5GIC and 6GIC, Institute for Communication Systems, University of Surrey, Guildford {\rm GU2 7XH}, United Kingdom}
\address[22]{ School of Information Science and Electronic Engineering, Shanghai Jiao Tong University, Shanghai {\rm 200240}, China}
\address[23]{ Guangdong Provincial Key Laboratory of Aerospace Communication and Networking Technology, HIT, Shenzhen {\rm 518055}, China}
\address[24]{ Future Network of Intelligent Institute, The Chinese University of Hong Kong (Shenzhen), Shenzhen {\rm 518066}, China\vspace*{-1em}}

\abstract{The International Telecommunication Union (ITU) identifies ``Artificial Intelligence (AI) and Communication” as one of six key usage scenarios for 6G. Agentic AI, characterized by its capabilities in multi-modal environmental sensing, complex task coordination, and continuous self-optimization, is anticipated to drive the evolution toward agent-based communication networks. Semantic communication (SemCom), in turn, has emerged as a transformative paradigm that offers task-oriented efficiency, enhanced reliability in complex environments, and dynamic adaptation in resource allocation. However, comprehensive reviews that trace their technological evolution in the contexts of agent communications remain scarce. Addressing this gap, this paper systematically explores the role of semantics in agent communication networks. We first propose a novel architecture for semantic-based agent communication networks, structured into three layers, four entities, and four stages. Three wireless agent network layers define the logical structure and organization of entity interactions: the intention extraction and understanding layer, the semantic encoding and processing layer, and the distributed autonomy and collaboration layer. Across these layers, four AI agent entities, namely embodied agents, communication agents, network agents, and application agents, coexist and perform distinct tasks. Furthermore, four operational stages of semantic-enhanced agentic AI systems, namely perception, memory, reasoning, and action, form a cognitive cycle guiding agent behavior. Based on the proposed architecture, we provide a comprehensive review of the state-of-the-art on how semantics enhance agent communication networks. Finally, we identify key challenges and present potential solutions to offer directional guidance for future research in this emerging field.
}
	\keywords{Semantic communication, agentic AI, AI agent, intellicise (intelligent and concise) wireless networks, communication and AI (ComAI)}
	
	\maketitle


\section{Introduction}

\subsection{Motivation}


The current fifth-generation (5G) mobile communication system exhibits limitations in coverage efficiency within a complex environment, cost and energy efficiency in the dense networking scenarios, and customization capabilities for vertical industries \cite{wang2023road}. These challenges have catalyzed the integration of Artificial Intelligence (AI) to address the interconnection demands of humans, machines, devices, and genies in future sixth-generation (6G) networks \cite{cui2025overview}.
In June 2023, the International Telecommunication Union-Radiocommunication (ITU-R) recommended that ``AI and Communication" be designated as one of six key usage scenarios for 6G. This underscores the principle that \textit{ubiquitous intelligence} will serve as a foundational design element across all 6G applications \cite{recommendation2023framework}.
On 13 February 2026, during the 131st RAN3 meeting of the 3rd Generation Partnership Project (3GPP) in Gothenburg, Sweden, 3GPP officially adopted ``aNB'' as the nomenclature for the 6G radio access network (RAN) node \cite{3gpp_R3_260548_2026}. Here, the prefix ``a'' signifies ``Advanced'' and ``AI'', signaling that the 6G base station (BS) will achieve breakthroughs in performance metrics and capabilities while fully integrating AI across critical domains. These include resource scheduling, channel modeling, network optimization, and maintenance, among other functional areas.

Agentic AI has recently garnered significant attention owing to its autonomous decision-making capabilities, driven by continuous perception-memory-reasoning-action loops \cite{sapkota2025ai,abou2025agentic}. Unlike conventional AI systems reliant on static inference pipelines \cite{fan2025generative,cheng2026apeg}, agentic AI integrates advanced generative frameworks, including large language models (LLMs), large vision models (LVMs), large multi-modal models (LMMs), and world models, to achieve multi-modal environmental sensing, complex task coordination, and continuous self-optimization \cite{zhang2026toward,jiang2026large}. These unique strengths make it well-suited for integration into the RAN, core network, and edge nodes of 6G networks, thereby advancing the evolution toward agent communication networks. The evolution from 1G to 6G is illustrated in Figure \ref{fig:The evolution from 1G to 6G}.
However, traditional bit-level transmission paradigms face critical challenges in supporting such networks, including substantial data transmission volumes, low interaction reliability, and excessive network resource consumption \cite{yang2022semantic}. In response, Semantic Communication (SemCom) has emerged as a transformative solution \cite{3gpp_TR22_870_v031_2025}, offering the following advantages:

\begin{figure}
    \centering
    \includegraphics[width=1\textwidth]{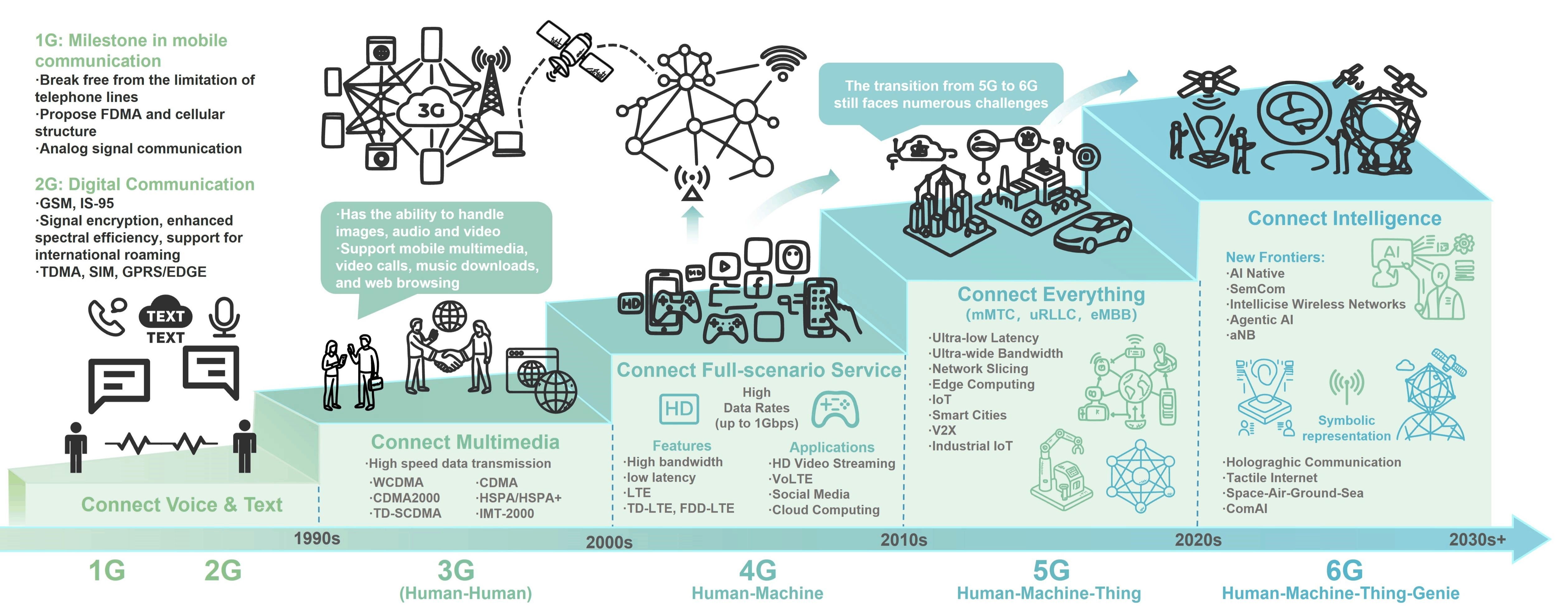} 
    \caption{The evolution from 1G to 6G.} 
    \label{fig:The evolution from 1G to 6G} 
\end{figure}

\begin{itemize}
\item \textbf{Task-oriented and Efficient Communication:} SemCom directly aligns with task objectives, minimizing spectrum and computational waste caused by redundant bit transmission \cite{xu2023task}. By extracting agent intentions, it supports real-time collaboration among massive agents, such as drone swarms and autonomous vehicle fleets \cite{gao2025agentic}.
\item \textbf{Highly Reliable Communication in Complex Environments:} 
SemCom reduces noise sensitivity and eliminates redundant information by transmitting core semantic features \cite{cao2026importance}. Additionally, through shared semantic knowledge bases (KBs), agents can reconstruct critical information via contextual reasoning, enhancing resilience to malicious interference and noise \cite{fan2025kgrag}.
\item \textbf{Flexible Resource Scheduling and Dynamic Adaptation:} 
SemCom dynamically adjusts transmission strategies based on semantic importance, enabling adaptive encoding and decoding in fluctuating environments and ensuring efficient network resource utilization \cite{meng2025semantic}.
\end{itemize}

Overall, SemCom alleviates data transmission burdens, enhances interaction reliability, and optimizes resource allocation efficiency for agent communication networks. Furthermore, SemCom-enabled \textit{intellicise (intelligent and concise)} wireless networks \cite{zhang2024intellicise,meng2026intellicise} integrate foundational theories, including information theory, AI theory, and system theory, to strengthen intention-driven, semantic-bearing, and distributed autonomy capabilities for agent communication networks. Moreover, the convergence of communication and AI (ComAI) \cite{zhang2025comai}, with SemCom and intellicise wireless networks as its core technologies, leverages grammatical, semantic, and pragmatic information through the integration of the user plane, control plane, computing plane, data plane, and intelligent plane, thereby advancing the evolution of agent communication networks.

To this end, the primary objective of this paper is to provide a comprehensive analysis of the rationale and mechanisms underlying SemCom integration within agent communication networks. 
We propose an architecture that consists of three wireless agent network layers, four AI agent entities, and four stages of agentic AI for semantic-based agent communication networks.
Furthermore, we conduct a systematic review of associated driving factors, key technologies, and their interdependencies, aiming to catalyze future research efforts in this rapidly evolving domain.

\subsection{Preliminaries and State-of-the-Art Works}

\subsubsection{Semantic Communications}
SemCom is an emerging paradigm that prioritizes the effective extraction, delivery, and interpretation of the underlying meaning of information, in contrast to conventional bit-level systems that focus on the accurate transmission of bit streams. By transmitting only task-relevant semantic features, SemCom achieves substantial gains in both efficiency and robustness. Moreover, it serves as a foundational enabler for intellicise wireless networks and ComAI. We summarize the relevant state-of-the-art survey, tutorial, and magazine literature in Table \ref{survey}.
In the context of general fundamentals and architectures, several comprehensive surveys and tutorials have systematically explored the theoretical foundations, early implementations, potential applications, and inherent challenges of semantics-empowered networks \cite{yang2022semantic, gunduz2022beyond, lu2023semantics, getu2025semantic}. More specifically, Chaccour \textit{et al.} \cite{chaccour2024less} propose the first rigorous end-to-end vision for semantic networks, while Shi \textit{et al.} \cite{shi2021semantic} introduce an architecture based on federated edge intelligence to enable resource-efficient semantic-aware networking.
Regarding security and resource management, works \cite{yang2024secure, meng2025survey, guo2024survey} provide comprehensive overviews and guidelines for designing secure SemCom systems, with an in-depth analysis of the interplay among network architectures, security paradigms, and privacy concerns. Additionally, Zhang \textit{et al.} \cite{zhang2025resource} present a systematic categorization and survey of advanced resource allocation strategies specifically tailored for SemCom networks.
In terms of advanced AI integration and the evolution toward native network intelligence, Liang \textit{et al.} \cite{liang2024generative} investigate the transformative role of generative AI architectures within SemCom systems. Building upon these foundations, \cite{zhang2024intellicise, meng2025semantic} offer in-depth studies on architectural innovations in intellicise wireless networks and semantic RANs. Finally, Meng \textit{et al.} \cite{meng2026intellicise} examine the emerging security and privacy implications at the intersection of intellicise networks and agentic AI.

\begin{table}[t!]
    \centering
    \renewcommand{\arraystretch}{1.1} 
    \scriptsize
    \setlength{\tabcolsep}{3pt}
    \caption{A classification of selected survey/tutorial/magazine papers}
    \label{survey}
    
    \begin{tabular}{@{}>{\centering\arraybackslash}m{0.10\textwidth}
                    >{\arraybackslash}m{0.05\textwidth}
                    >{\arraybackslash}m{0.05\textwidth}
                    >{\arraybackslash}m{0.08\textwidth}
                    >{\arraybackslash}m{0.65\textwidth}@{}}
        \toprule
        \textbf{Domain} & \textbf{Ref.} & \textbf{Year} & \textbf{Type} & \textbf{Contributions}\\
        \midrule    
        \multirow{14}{=}[-18ex]{\centering{SemCom}}  & \cite{shi2021semantic} & 2021 & Magazine & Propose an architecture based on federated edge intelligence for supporting resource-efficient semantic-aware networking. \\
        \cmidrule{2-5}
         & \cite{yang2022semantic} & 2022 & Survey & Systematically investigate the fundamentals, potential applications, and intrinsic challenges of SemComs for the future internet. \\
        \cmidrule{2-5}
         & \cite{gunduz2022beyond} & 2023 & Tutorial & Summarize from the early adaptation, semantic perception, and task-oriented communication, covering the basics, algorithms, and potential implementations. \\
        \cmidrule{2-5}
         & \cite{lu2023semantics} & 2024 & Tutorial & Deliver a comprehensive tutorial-cum-survey detailing the landscape of semantics-empowered communication systems. \\
        \cmidrule{2-5}
         & \cite{yang2024secure} & 2024 & Magazine & Provide a comprehensive guide on how to design secure SemCom systems in the real-world wireless communication network. \\
        \cmidrule{2-5}
         & \cite{zhang2024intellicise} & 2025 & Survey & Provide an in-depth research of intellicise wireless networks derived from SemComs. \\
        \cmidrule{2-5}
         & \cite{liang2024generative} & 2025 & Survey & Explore the transformative integration of Generative AI architectures and technologies within SemCom networks. \\
        \cmidrule{2-5}  
         & \cite{meng2025survey} & 2025 & Survey & Provide a comprehensive overview of the techniques that can be used to ensure the security of SemCom. \\
        \cmidrule{2-5}
         & \cite{chaccour2024less} & 2025 & Survey & Present the first rigorous and holistic end-to-end vision of SemCom networks. \\
        \cmidrule{2-5}
         & \cite{getu2025semantic} & 2025 & Survey & Introduce the theoretical foundation, research environment, market environment, opportunities and challenges, and future research directions of SemCom. \\
        \cmidrule{2-5}   
         & \cite{guo2024survey} & 2025 & Survey & Analyze the intersection of network architecture, security paradigms, and privacy issues in SemCom environments. \\
        \cmidrule{2-5}
         & \cite{zhang2025resource} & 2026 & Survey & Systematically categorize and surveys advanced resource allocation strategies tailored for wireless SemCom networks. \\
        \cmidrule{2-5} 
         & \cite{meng2026intellicise} & 2026 & Magazine & Investigate the profound security and privacy implications benefiting from the convergence of intellicise networks and Agentic AI. \\
       \cmidrule{2-5}
         & \cite{meng2025semantic} & 2026 & Survey & Discusse the state-of-the-art architectural innovations and future evolutionary paths of semantic RANs. \\
        \midrule
        \multirow{7}{=}[-10ex]{\centering{Agentic AI}} & \cite{acharya2025agentic} & 2025 & Survey & Offer a holistic survey on autonomous intelligence, focusing on the deployment of Agentic AI for resolving complex objectives. \\
        \cmidrule{2-5}
         & \cite{sapkota2025ai} & 2025 & Survey & Establish a rigorous conceptual taxonomy distinguishing AI agents from Agentic AI, alongside an analysis of their applications. \\
        \cmidrule{2-5}
         & \cite{datta2025agentic} & 2025 & Survey & Critically introduce and evaluate the threat landscapes, defense mechanisms, and open security challenges inherent in Agentic AI systems. \\
        \cmidrule{2-5}
         & \cite{abou2025agentic} & 2025 & Survey & Surveys the diverse architectural paradigms, practical applications, and future trajectories of Agentic AI technologies. \\
        \cmidrule{2-5}
         & \cite{wang2025ai} & 2025 & Survey & Reviews the underpinning techniques, developmental challenges, and emerging opportunities in AI agentic programming. \\
        \cmidrule{2-5}
         & \cite{jiang2026large} & 2026 & Tutorial & Systematically and comprehensively introduce the principle, design and application of large AI models (LAMs) in intelligent communication systems.  \\
       \cmidrule{2-5}
         & \cite{zhang2026toward} & 2026 & Survey &  Provide a comprehensive survey of Agentic AI and agentification frameworks tailored for edge general-purpose intelligence. \\
         \bottomrule
    \end{tabular}
\end{table}

\renewcommand{\arraystretch}{1.0}
\setlength{\tabcolsep}{6pt}
\normalsize
\subsubsection{Agentic AI Systems}
Agentic AI refers to highly autonomous AI systems capable of advanced cognitive reasoning, environmental perception, and independent decision-making. Unlike traditional passive AI systems or foundational LAMs, agentic AI distinguishes itself through continuous interaction with dynamic environments and the autonomous execution of multi-step, complex goals via structured perception-action loops. We summarize recent survey and tutorial literature on agentic AI in Table \ref{survey}.
To clarify core concepts in this rapidly evolving field, Sapkota \textit{et al.} \cite{sapkota2025ai} establish a rigorous taxonomy that distinguishes agentic AI from conventional AI agents.
Delving into specific technical dimensions, Wang \textit{et al.} \cite{wang2025ai} review the underlying techniques, developmental challenges, and emerging opportunities in agentic AI programming. From a security standpoint, Datta \textit{et al.} \cite{datta2025agentic} critically examine threat landscapes, defense mechanisms, and unresolved security challenges inherent in agentic AI systems.
In the telecommunications domain, the integration of agentic AI is gaining increasing relevance. Jiang \textit{et al.} \cite{jiang2026large} provide a systematic introduction to the designs of LAMs and agentic AI technologies within intelligent communication systems. Finally, Zhang \textit{et al.} \cite{zhang2026toward} present a comprehensive survey on agentic AI and agentification frameworks specifically designed to enable edge general-purpose intelligence.

\subsection{Key Contributions and Outline}

Although numerous researchers have focused on SemCom \cite{meng2025image,shi2021semantic,yang2022semantic,gunduz2022beyond,lu2023semantics,yang2024secure,zhang2024intellicise,liang2024generative,meng2025survey,chaccour2024less,getu2025semantic,guo2024survey,zhang2025resource,meng2026intellicise,meng2025semantic} and agentic AI \cite{acharya2025agentic,sapkota2025ai,datta2025agentic,abou2025agentic,wang2025ai,jiang2026large,zhang2026toward}, a comprehensive understanding of the state-of-the-art in semantic-based agent communication networks remains nascent. For instance, Gao \textit{et al.} \cite{gao2025agentic} propose a unified agentic AI-enhanced SemCom framework, while Yu \textit{et al.} \cite{yu2025semantic} develop a semantic-driven AI agent communication framework. Notably, the literature currently lacks a systematic review tracing the evolutionary trajectory of both semantic-based agentic AI systems and semantic-based agent wireless networks. To address this gap, this survey presents a comprehensive exploration of state-of-the-art technologies for semantic-based agent communication networks. The main contributions are summarized as follows.

\subsubsection{Outline the Architecture of Semantic-based Agent Communication Networks}
We propose a novel architecture that consists of three layers, four entities, and four stages for semantic-based agent communication networks, where the components interact to form a closed-loop system. The architecture comprises three wireless agent network layers that define the logical structure and organization of entity interactions: the intention extraction and understanding layer, the semantic encoding and processing layer, and the distributed autonomy and collaboration layer. Four AI agent entities, including embodied agent, communication agent, network agent, and application agent, inhabit the network and execute tasks. Finally, four operational stages for semantic-enhanced agentic AI systems, including perception, memory, reasoning, and action, define the cognitive cycle that guides agent behavior.

\subsubsection{Explore the State-of-the-Art in Semantic-based Agent Communication Networks}
Based on the proposed architecture, we investigate recent advancements in the role of semantics in enhancing agent communication networks. 
\begin{figure}[t!]
    \centering
    \includegraphics[width=0.7\linewidth]{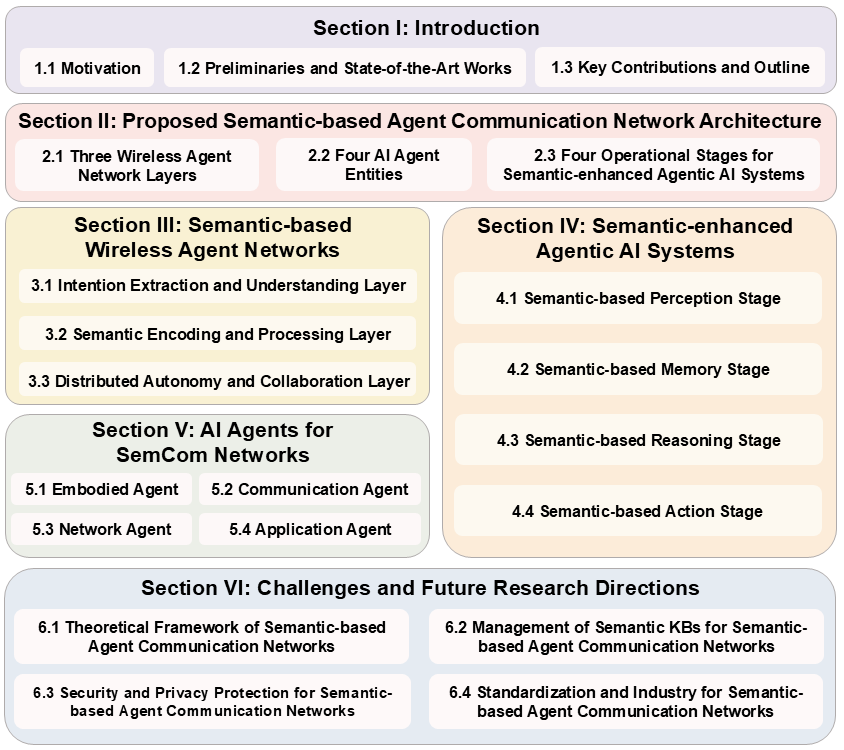}
    \caption{The structure of this paper.}
    \label{fig:structure}
\end{figure}

\begin{itemize}
\item \textit{Three Layers:} For the intention extraction and understanding layer, we review representative approaches, including evidence-based intent inference, opponent modeling-based intent extraction, and mind modeling-based intent inference. For the semantic encoding and processing layer, we summarize representative approaches, including semantic-based coding, semantic-based beam management, semantic-based Channel State Information (CSI) feedback, semantic-based Hybrid Automatic Repeat request (HARQ), Age of Semantic Information (AoSI), and semantic Knowledge Base (KB). For the distributed autonomy and collaboration layer, we explore representative methods, including distributed access, knowledge collaboration, and resource scheduling. 
\item \textit{Four Stages:} For the semantic-based perception stage, we summarize representative techniques, including semantic feature extraction and representation, task-oriented environmental sensing, semantic object grounding and tracking, and embodied semantic environment understanding. For the semantic-based memory stage, we examine representative approaches, including hierarchical semantic memory structure, semantic retrieval and reasoning, memory evolution and knowledge update, and cognitive augmentation via memory. For the semantic-based reasoning stage, we review representative methods, including chain-of-thought (CoT) reasoning, knowledge graph (KG)-augmented reasoning, retrieval-augmented reasoning, tree-structured multi-path reasoning, and neuro-symbolic reasoning. For the semantic-based action stage, we summarize representative approaches, including semantic tool acquisition, reasoning-action interleaving, multi-agent collaborative action, semantic self-correction, and reinforcement-based semantic feedback.
\item \textit{Four Entities:} For embodied agents, we highlight two representative projects: SayCan and Atlas. For communication agents, we review the semantic-driven AI agent communication framework and agentic AI-enhanced SemCom framework, followed by an introduction to two notable implementations, Channel GPT and UniClaw. For network agents, we present four key projects, RAN Agent, Agentic AI for RAN, JoinAI-Agent, and Xingchen Super Agent. For application agents, we provide an overview of two representative general agents, along with several vertical agents tailored for smart factory, smart healthcare, smart city, and intelligent transportation.
\end{itemize}
\subsubsection{Discuss Challenges and Potential Solutions}
While extensive research has explored the use of semantics to enhance agent communications, several fundamental challenges remain. Building on this analysis, we systematically identify these challenges and propose potential research directions. Key areas of focus include the theoretical framework, the management of semantic KBs, security and privacy protection, and standardization and industry adoption in semantic-based agent communication networks.

\textit{Roadmap:} The outline of this review is depicted in Figure \ref{fig:structure}.
Specifically, Section \ref{sectionarchitecture} presents the architecture of semantic-based agent communication networks. Section \ref{sectioncommunication} reviews key technologies supporting the three layers of semantic-based wireless agent networks. Section \ref{sectionagenticai} discusses how semantics enhance the four stages of agentic AI systems. Section \ref{sectionagent} examines four AI agent entities for SemCom networks. Section \ref{sectionfuture} outlines challenges and potential directions. Finally, Section \ref{sectionconclusion} concludes this review.


\section{Proposed Semantic-based Agent Communication Network Architecture}
\label{sectionarchitecture}

\begin{figure}
    \centering
    \includegraphics[width=0.9\textwidth]{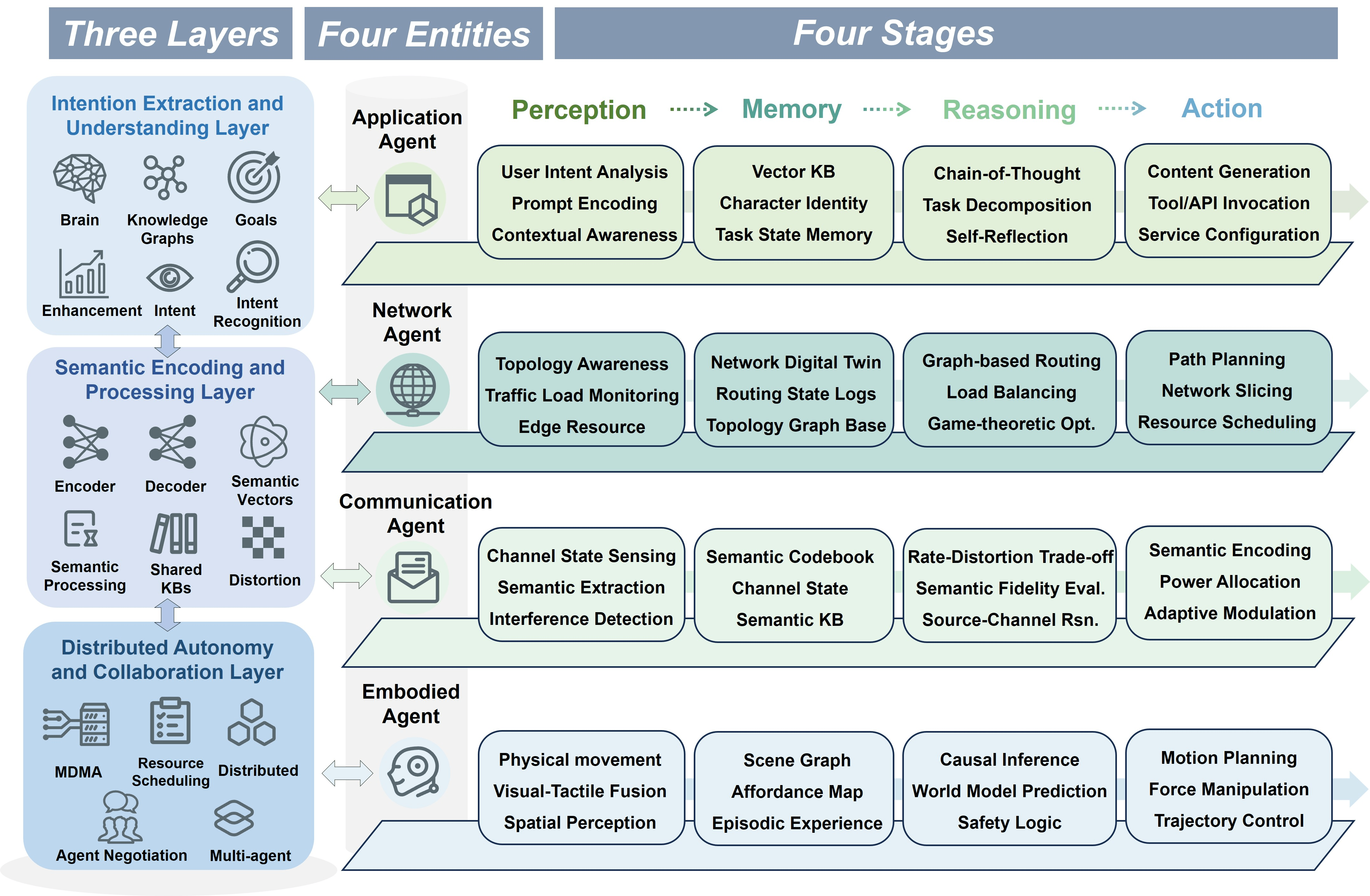} 
    \caption{The proposed architecture for semantic-based agent communication networks, comprising three layers, four entities, and four stages.} 
    \label{fig:networkarchitecture} 
\end{figure}

As illustrated in Figure \ref{fig:networkarchitecture}, we propose an architecture for semantic-based agent communication networks, comprising three layers, four entities, and four stages.
\begin{itemize}
\item 
\textbf{``Three Layers'':} Define the agent communication system's logical organizational framework through three core layers: intention extraction and understanding layer, semantic encoding and processing layer, and distributed autonomy and collaboration layer.
\item 
\textbf{``Four Entities'':} Explicitly categorize communication network entities into four distinct AI agent types: embodied agents, communication agents, network agents, and application agents, each representing a unique physical and functional role within the architecture.
\item 
\textbf{``Four Stages'':} Outline the semantic-enhanced workflow for AI agents across four sequential stages: perception, memory, reasoning, and action, capturing the end-to-end cognitive processing pipeline.
\end{itemize}

The relationship between them is as follows. Firstly, three layers define the overarching structure of the communication network, establishing how the intelligent network system is organized and determining where four entities operate within it. Secondly, four entities serve as the functional carriers, identifying task executors and driving the execution of four stages. Thirdly, four stages outline the behavioral methodology, addressing the question of how agents think. This process occurs not only within individual agents but also across three layers, thereby creating a closed loop for information flow.

\subsection{Three Wireless Agent Network Layers}

Table \ref{3layer} provides a concise comparison of these three layers in terms of their core focuses and operational levels, with detailed explanations provided below.

\begin{itemize}
\item 
\textbf{Intention Extraction and Understanding Layer:}
In traditional communication systems, the control plane is primarily concerned with how data is forwarded. The intention layer, by contrast, introduces two elements that have never been addressed in traditional communication: knowledge and goals. Knowledge provides it with the capacity to understand the world, going beyond mere sensing \cite{reily2022real}; goals enable autonomous decision-making, rather than simply reacting to inputs \cite{sohrabi2016plan}. Specifically, the intention layer perceives raw data streams from various sensors \cite{wang2021tom2c}, such as camera pixels, LiDAR point clouds, and inertial measurement unit data, transforms these signals into structured knowledge \cite{reily2022real}, and fuses them to resolve ambiguity \cite{shi2025muma}. By leveraging pre-deployed KBs such as KGs and world models, the intention layer forms a contextualized understanding of current perceptual information, and performs intent recognition. Based on this understanding, it defines communication goals that specify the intended recipients, the knowledge content to be conveyed, and the desired effects.
For example, when an autonomous vehicle detects a traffic accident 50 meters ahead, it recognizes that this newly acquired information is critical for surrounding vehicles. It then generates a communication goal: ``Broadcast to vehicles within a 300-meter radius behind: `accident ahead, slow down and detour'." Here, the recipients are vehicles 300 meters to the rear, the knowledge content includes the location and severity of the accident, and the desired effect is for those vehicles to slow down or alter their route.

\item 
\textbf{Semantic Encoding and Processing Layer:}
This layer aims to simplify information representation and ensure communication robustness, all while preserving the original meaning. It leverages a shared semantic KB that serves as prior knowledge for both the transmitting and receiving agents \cite{fan2025kgrag}. Semantic KB is typically aligned through pre-training techniques prior to communication and is dynamically updated during operation. Upon receiving a communication task from the intent layer, the transmitting agent employs a semantic encoder to extract semantic information from both the source and the channel, encoding it into a semantic vector. The receiving agent then performs the reverse operation to fulfill the communication task. Whereas traditional communication relies on bit error rate (BER) as its performance metric, the semantic layer emphasizes semantic distortion rate (SDR), which measures the discrepancy between the meaning reconstructed by the receiver and the original intent of the sender \cite{niu2025mathematical}. The semantic layer can tolerate a certain degree of physical-layer bit errors, provided that these errors do not compromise the core meaning.

\item 
\textbf{Distributed Autonomy and Collaboration Layer:}
The objective of this layer is to enable efficient, reliable, and low-latency collaboration among multiple agents operating in highly dynamic and complex wireless environments. Through local perception and autonomous negotiation, agents coordinate communication tasks in a distributed manner.
Firstly, the layer facilitates efficient multi-user access via Model Division Multiple Access (MDMA) \cite{zhang2023model,liang2024orthogonal,cao2026s}. MDMA extracts high-dimensional source features by adopting a model-driven approach grounded in semantics, thereby constructing a model-based information space that accounts for the characteristics of both multi-modal sources and communication channels. MDMA differentiates users based on the semantic features embedded within the models. Secondly, this layer employs semantic-aware resource scheduling strategies that prioritize resource allocation for high-priority or time-sensitive communication tasks \cite{zhang2025resource}. Finally, agents with complementary perceptual capabilities, computational resources, and knowledge backgrounds collaborate to overcome individual limitations, thereby enabling the execution of complex semantic understanding and communication tasks \cite{razlighi2025cooperative}.

\begin{table}
    \centering
    \caption{Three layers of semantic-based wireless agent networks}
    \label{3layer}
	\renewcommand{\arraystretch}{1.5}
	\footnotesize
    \begin{tabular}{>{\arraybackslash}m{0.4\textwidth} >{\arraybackslash}m{0.20\textwidth} >{\arraybackslash}m{0.2\textwidth}}
        \toprule
        \textbf{Logical Layers} & \textbf{Addressed Issues} & \textbf{Levels} \\
        \midrule
        Intention Extraction and Understanding Layer  & Why Communication? & Knowledge and Goal \\
        Semantic Encoding and Processing Layer  & What to Transmit? & Meaning and Significance \\
        Distributed Autonomy and Collaboration Layer  & How to Communicate? & Node and Flow \\
        \bottomrule
    \end{tabular}
\end{table}

\end{itemize}

\subsection{Four AI Agent Entities}

\begin{table}
    \centering
    \caption{Four Entities of AI agents in SemCom networks}
    \label{4entity}
	\renewcommand{\arraystretch}{1.5}
	\footnotesize
    \begin{tabular}{>{\arraybackslash}m{0.18\textwidth} >{\arraybackslash}m{0.18\textwidth} >{\arraybackslash}m{0.35\textwidth}}
        \toprule
        \textbf{Agent Entities} & \textbf{Roles} & \textbf{Main Functions} \\
        \midrule
        Embodied Agents  & Physical Execution & Implementation and Feedback of Semantics \\
        Communication Agents  & Interactive Connection & Coding and Transmission of Semantics \\
        Network Agents  & Network Management & Scheduling and Guarantee of Semantics \\
        Application Agents  & Service Provision & Generation and Consumption of Semantics \\
        \bottomrule
    \end{tabular}
\end{table}

Table \ref{4entity} offers a concise comparison of these four AI agent entities, highlighting their respective roles and primary functions. A more detailed account of each is provided below.

\begin{itemize}
\item 
\textbf{Embodied Agent:}
Embodied agents are active participants in the physical world, capable of understanding, reasoning, and interacting with their surroundings in real time\cite{duan2022survey}. They typically take the form of various robotic systems, such as humanoid robots, drones, and autonomous vehicles. While executing tasks, these agents must transmit multi-modal data in real time, such as visual and tactile information. SemCom enables them to transmit only environmental changes or target features instead of full video streams, significantly reducing the burden on air interface resources. Moreover, because tactile and visual perceptions are inherently correlated in the physical world, SemCom facilitates the semantic alignment of information across different modalities, leading to faster physical responses.

\item 
\textbf{Communication Agent:}
Communication agents serve as the essential link between other agents and the capabilities of SemCom networks, allowing agents to natively access and utilize core network functions. Specifically, these agents encapsulate the underlying semantic encoding and decoding capabilities. When an embodied or application agent needs to transmit messages, the communication agent generates an encoded semantic vector at the transmitter and reconstructs the original meaning through the reverse process at the receiver\cite{gao2025agentic}. Moreover, since different agents may operate with distinct KBs, communication agents engage in protocol-based negotiation to establish a  consistent semantic background, thereby enabling effective SemCom links. Additionally, they are responsible for the dynamic updating and synchronization of semantic KBs, which is crucial for the ongoing enhancement of intelligent communication capabilities.

\item 
\textbf{Network Agent:}
The primary purpose of network agents is to integrate agentic AI technologies into the network infrastructure, enabling self-intelligence\cite{demirel2026intents}. While traditional networks merely schedule data flows, network agents harness their understanding and reasoning capabilities to orchestrate semantic routing. They perform in-network aggregation, optimization, and intelligent distribution, thereby achieving efficient semantic aggregation. Network agents support intent-based network management by automatically translating high-level operator intentions into specific configurations, facilitating network self-configuration, self-optimization, and self-healing. Furthermore, network agents can rapidly locate and diagnose faults, conduct pre-deployment network simulations to mitigate risks, accurately identify congestion to optimize service quality, and analyze data across disparate systems and domains to proactively detect anomalies.

\item 
\textbf{Application Agent:}
Application agents primarily deliver customized information and services to users, operating on end devices or at the network edge. These agents support a wide range of industries, such as smart factories, intelligent healthcare, and smart cities\cite{abou2025agentic}. By utilizing pre-trained semantic KBs and generating personalized semantic KGs for users, they enhance both intent extraction and semantic representation. Furthermore, the inference processes of LAMs demand significant computational resources. To address this, resource-constrained application agents can offload complex semantic understanding tasks to the network side, focusing locally only on semantic presentation and lightweight data collection.

\end{itemize}

\subsection{Four Operational Stages for Semantic-enhanced Agentic AI Systems}

\begin{table}
    \centering
    \caption{Four stages of semantic-enhanced agentic AI systems}
    \label{4stage}
	\renewcommand{\arraystretch}{1.5}
	\footnotesize
    \begin{tabular}{>{\arraybackslash}m{0.08\textwidth} >{\arraybackslash}m{0.35\textwidth} >{\arraybackslash}m{0.28\textwidth}}
        \toprule
        \textbf{Stages} & \textbf{Main Functions} & \textbf{Enhancement of Semantics} \\
        \midrule
        Perception  & Information Acquisition and Representation & Improved Matching and Alignment \\
        Memory  & Knowledge Storage and Management & Stronger Association and Retrieval \\
        Reasoning  & Decomposition of Tasks & Deeper Reasoning and Analogy \\
        Action  & Execution of Instructions & Faster Execution and Feedback \\
        \bottomrule
    \end{tabular}
\end{table}

Table \ref{4stage} summarizes the four stages by outlining their core functions and how semantics enhance them. Further elaboration is provided as follows.

\begin{itemize}
\item 
\textbf{Perception Stage:}
The agent perceives multi-modal information from the environment and converts unstructured data into a structured representation it can process \cite{miao2023occdepth,rosinol20203d}. Traditional perception techniques can identify content but often fail to grasp the relationships between disparate pieces of information or the underlying intent within a specific context. In contrast, semantic-enhanced perception integrates multi-modal data into a unified semantic space, achieving semantic alignment \cite{girdhar2023imagebind,liu2024grounding}. This enables the agent to map vague user instructions to concrete, actionable intentions.

\item 
\textbf{Memory Stage:}
The agent stores and manages knowledge, typically divided into short-term memory and long-term memory. The former handles ongoing conversations or tasks \cite{bulatov2022recurrent,park2023generative}, while the latter preserves historical experiences, common sense, and learned knowledge \cite{wang2023augmenting,shinn2023reflexion}. 
Traditional databases, which primarily store keywords, make it difficult for conventional memory systems to link successive instructions. This limitation hinders an agent's ability to accumulate experience and provide continuous, personalized service. Semantics, however, enable the construction of KGs that connect disparate memory fragments into a cohesive network \cite{park2023generative,chhikara2025mem0}. When a new task arises, semantic-enhanced memory can retrieve relevant information based on conceptual similarity, allowing for more accurate association and retrieval \cite{sanmartin2024kg,asai2023self}.

\item 
\textbf{Reasoning Stage:}
The agent performs analysis, planning, decision-making, and logical deduction based on perceived information and stored knowledge. Traditional symbolic reasoning often struggles with the ambiguity and common-sense nuances of the real world. By leveraging large-scale semantic KBs, an agent can be equipped with common-sense knowledge during its analysis and planning phases \cite{jiang2023structgpt}. Furthermore, through semantic similarity, the agent can transfer solutions from previously solved problems to novel ones, enabling analogical reasoning \cite{lightman2024}. When handling complex tasks, semantics help ensure coherence in the chain of thought through multi-step logical inference.

\item 
\textbf{Action Stage:}
The agent translates the decisions derived from reasoning into concrete actions. Traditional action techniques are heavily dependent on strict application programming interface (API) formats, where even slight deviations in instruction can cause the system to fail. In contrast, a semantic-enhanced agent comprehends user intent and automatically maps it to the required API parameters \cite{park2023generative}. Moreover, if an error occurs during action, semantic analysis can diagnose the cause and automatically adjust the execution strategy in real time \cite{shinn2023reflexion,madaan2023selfrefine}.

\end{itemize}

\section{Semantic-based Wireless Agent Networks}
\label{sectioncommunication}

This section provides an overview of the technologies that empower semantic-based wireless agent networks, spanning the layers of intention extraction and understanding, semantic encoding and processing, and distributed autonomy and collaboration.

\subsection{Intention Extraction and Understanding Layer}

The intention extraction and understanding layer enables agents to interpret goals and intentions, thereby facilitating coordinated decision-making. As illustrated in Table \ref{Intention Extraction and Understanding}, we review several representative approaches within this layer, including evidence-based intent inference, opponent modeling-based intent extraction, and mind modeling-based intent inference.

\begin{table}
    \centering
    \caption{Representative approaches for the intention extraction and understanding layer}
    \label{Intention Extraction and Understanding}
	\renewcommand{\arraystretch}{1.5}
	\footnotesize
    \begin{tabular}{>{\arraybackslash}m{0.13\textwidth} >{\arraybackslash}m{0.23\textwidth} >{\arraybackslash}m{0.56\textwidth}}
        \toprule
        \textbf{Category} & \textbf{Sub-Category} & \textbf{Descriptions} \\
        \midrule
        \multirow{4}{=}{Evidence-based Intent Inference}  
        & Behavior-based Recognition & Infers agent goals by matching observed action sequences to learned or structured behavioral patterns \cite{dann2023multi,su2023fast}. \\
        \cmidrule{2-3}
        & Plan Recognition as Planning & Uses planning algorithms to generate candidate plans and infer goals from their consistency with observations. \cite{ramirez2009plan,sohrabi2016plan,shvo2018ai,masters2019goal}. \\
        \cmidrule{2-3}
        & Landmark-based Recognition & Identifies landmarks in plans to infer the most probable goal by matching behaviors to those milestones \cite{pereira2020landmark,wilken2023planning,wilken2023leveraging,zhang2023intention}. \\
        \cmidrule{2-3}
        & Active Goal Recognition & Strategically gathers information by influencing observation through its own actions, enhancing goal inference accuracy \cite{zhang2025probabilistic,shvo2020active,maisto2023interactive}. \\
        \midrule
        \multirow{5}{=}{Opponent Modeling-based Intent Inference}
        & Hypothesis-based Modeling & Uses hypothesised agent types to match observed actions and update beliefs about the agent's behavior \cite{albrecht2016belief,albrecht2019reasoning,zhu2025single}. \\
        \cmidrule{2-3}
        & Subgoal-based Modeling & Models opponent by analyzing subgoals to better generalize to unknown opponents \cite{yu2024opponent}. \\
        \cmidrule{2-3}
        & Reward Inference & Estimates opponent goals by inferring their reward function through inverse RL \cite{lin2019multi,fu2021evaluating,freihaut2024feasible}. \\
        \cmidrule{2-3}
        & Team Modeling & Infers team goals by analyzing the collective actions of agents within the team \cite{ying2023inferring,reily2022real}. \\
        \cmidrule{2-3}
        & Policy-based Modeling & Uses policy-level reasoning over observed actions to infer hidden goals or evolving strategies \cite{raileanu2018modeling,zhang2020opponent,yu2022model}. \\
        \midrule
        \multirow{3}{=}{Mind Modeling-based Intent Inference}
        & Bayesian Mind Models & Uses probabilistic models to estimate agents’ mental states and intentions, managing uncertainty in observed behaviors \cite{poppel2018satisficing,lim2020improving}. \\
        \cmidrule{2-3}
        & Mental State Modeling & Models agents’ mental states by analyzing observed behaviors, enabling predictions of their intentions \cite{rabinowitz2018machine,wang2021tom2c,shi2025muma}. \\
        \cmidrule{2-3}
        & LLM‑based Mind Modeling & Uses LLMs to model agents' internal mental states and intentions \cite{cross2024hypothetical,li2023theory,li2025theory}. \\
        \bottomrule
    \end{tabular}
\end{table}

\subsubsection{Evidence-based Intent Inference}

Evidence-based intent inference derives agents’ intentions directly from observable evidence such as actions, trajectories, and interaction patterns. These approaches focus on identifying behavioral patterns and associating them with possible goals or intentions. By leveraging observable signals, they offer a practical means of inferring intentions in multi-agent environments. Representative methods include behavior-based recognition, plan recognition as planning, landmark-based recognition, and active goal recognition.

\paragraph{Behavior-based Recognition}

\begin{figure}
    \centering
    \includegraphics[width=0.8\linewidth]{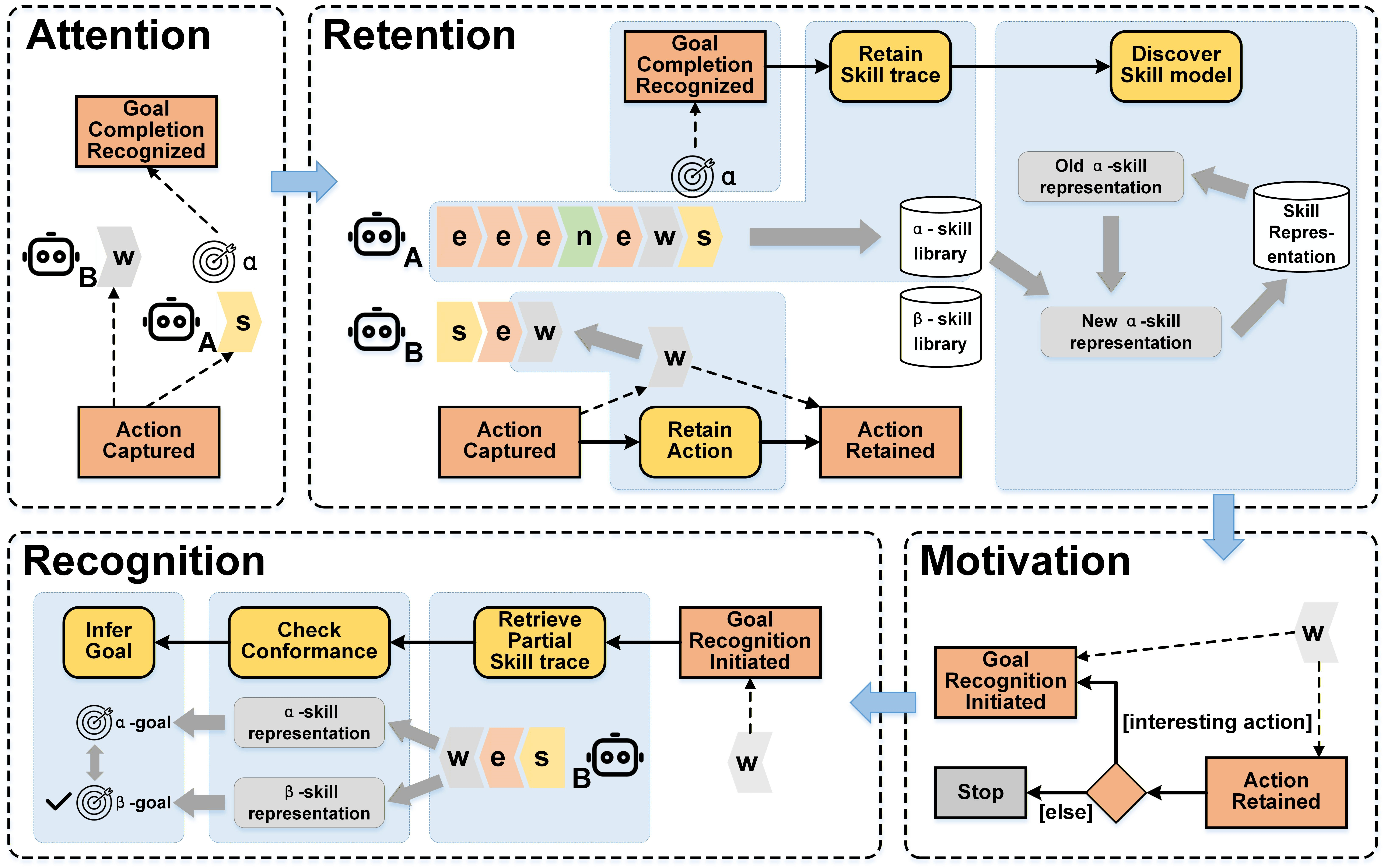}
    \caption{Illustration of the continuous goal recognition framework \cite{su2023fast}, consisting of the attention, retention, motivation, and recognition stages for capturing relevant actions, retaining skill traces and representations, triggering recognition, and inferring goals.}
    \label{fig:Intention}
\end{figure}

Behavior-based recognition infers agents’ intentions by analyzing observable action sequences and matching them to learned or structured behavioral patterns. Dann \textit{et al.} \cite{dann2023multi} propose a multi-agent intention recognition framework that employs online goal recognition to infer other agents’ goals from their behaviors and subsequently predict their likely future actions. This method supports rapid recognition even when agents pursue multiple goals in parallel. As illustrated in Figure \ref{fig:Intention}, Su \textit{et al.} \cite{su2023fast} introduce a data-driven goal recognition framework that learns skill representations from observed behavioral histories. The framework infers goals by analyzing discrepancies between ongoing behaviors and these learned  representations, and remains effective even without complete environmental knowledge or prior information about agents.

\paragraph{Plan Recognition as Planning}

Plan recognition as planning frames intention inference as a planning problem, using planning algorithms to generate or evaluate plans for candidate goals and identify the goals that best explain the observations \cite{ramirez2009plan,masters2019goal}. Sohrabi \textit{et al.} \cite{sohrabi2016plan} enhance this approach by transforming the recognition task into a planning problem with action costs, introducing additional explain and discard actions to manage unreliable observations, and defining posterior probabilities over plans and goals. This method significantly improves recognition performance in the presence of noisy observations. Additionally, Shvo \textit{et al.} \cite{shvo2018ai} propose a planning-based method for multi-agent plan recognition by compiling the recognition task into a temporal planning problem with durative actions, making it suitable for settings involving temporal actions and potentially unreliable observations.

\paragraph{Landmark-based Recognition}

Landmark-based recognition infers intentions by identifying critical intermediate states or milestones that must be achieved for specific goals \cite{pereira2020landmark,wilken2023planning,zhang2023intention}. It matches observed behaviors with these landmarks to recognize intentions. Pereira \textit{et al.} \cite{pereira2020landmark} propose a landmark-based goal recognition method that computes planning landmarks for candidate goals and uses them to perform recognition. This method further estimates goal completion by measuring the proportion of achieved landmarks. It improves recognition speed by approximately 8.6 times compared to prior methods while maintaining accuracy. Wilken \textit{et al.} \cite{wilken2023leveraging} introduce a landmark-based hybrid recognition method that replaces the planning-based component in mainstream hybrid goal recognition with a landmark-based alternative. This approach significantly reduces recognition time, enables rapid goal recognition in complex scenarios, and enhances overall performance.

\paragraph{Active Goal Recognition}

Active goal recognition enhances goal inference by enabling the observer to actively gather information through its own actions, thereby reducing uncertainty and accelerating the recognition process \cite{shvo2020active,maisto2023interactive}. Shvo \textit{et al.} \cite{shvo2020active} propose an active goal recognition approach in which the observer can sense, act, or interact with the environment to obtain more informative observations. This method facilitates earlier recognition and can make goal identification possible even when passive observation alone is insufficient. Zhang \textit{et al.} \cite{zhang2025probabilistic} present a probabilistic active goal recognition framework that enables the observer to select informative actions during recognition. By updating goal beliefs under uncertainty, this method improves goal disambiguation and recognition accuracy.

\subsubsection{Opponent Modeling-based Intent Inference}

Compared with evidence-based approaches that primarily rely on observed behaviors, opponent modeling-based intent inference involves explicitly constructing models of other agents. By capturing opponents’ strategies or behavioral tendencies, these methods enable more predictive reasoning about their goals and intentions in interactive environments. Representative methods include hypothesis-based modeling, subgoal-based modeling, reward inference, team modeling, and policy-based modeling.

\paragraph{Hypothesis-based Modeling}

Hypothesis-based modeling explicitly maintains multiple hypotheses about the possible behaviors of other agents and infers intentions by updating their likelihoods based on observed actions \cite{albrecht2016belief,albrecht2019reasoning}. Albrecht \textit{et al.} \cite{albrecht2016belief} propose a hypothesis-based method that represents possible opponent behaviors as hypothesis types and updates posterior beliefs over them using observed actions. Zhu \textit{et al.} \cite{zhu2025single} introduce a unified type-based framework that maintains multiple hypothetical policies for previously unknown opponents and updates beliefs over them from observed actions. This framework captures other agents’ behavioral tendencies through these evolving hypotheses.

\paragraph{Subgoal-based Modeling}

Subgoal-based Modeling infers intentions by modeling the intermediate subgoals underlying opponents’ behaviors. Since agents with different strategies may share the same subgoals, this approach enables more generalizable opponent understanding. Yu \textit{et al.} \cite{yu2024opponent} propose an opponent modeling method based on subgoal inference, which predicts opponents’ future subgoals from historical trajectories and uses them to characterize behavioral tendencies. Compared with action-based opponent modeling, this method improves adaptation to unknown opponents. In scenarios involving collaboration with unknown opponents, it achieves a 5\% to 20\% improvement in task success rate over previous approaches.

\paragraph{Reward Inference}

Reward inference infers intentions by estimating the reward functions that best explain opponents’ observed behaviors in multi-agent interactions \cite{lin2019multi,freihaut2024feasible}. Lin \textit{et al.} \cite{lin2019multi} propose a multi-agent inverse reinforcement learning (RL) method for general-sum stochastic games, which infers players’ rewards under different equilibrium assumptions by solving constrained optimization problems. Fu \textit{et al.} \cite{fu2021evaluating} introduce a scalable multi-agent inverse RL approach that reduces the multi-agent problem to a set of single-agent inverse learning problems while preserving strong rationality.

\paragraph{Team Modeling}

Team modeling infers intentions by modeling the collective behavior and goals of a team rather than individual agents. By capturing the interactions and coordination among team members, it provides a comprehensive understanding of team intent in cooperative environments. Reily \textit{et al.} \cite{reily2022real} propose a method for real-time team behavior recognition using multisensory data. The approach embeds team behaviors into a graph structure and uses robot learning to model individual actions and interrelationships, enabling real-time recognition of team goals and improving cooperation in dynamic environments. Ying \textit{et al.} \cite{ying2023inferring} model a cooperative team as a single collective agent to simplify inference. By analyzing both actions and instructions, this scheme infers the shared goals of the team and enhances understanding of team-level behaviors in cooperative environments.

\paragraph{Policy-based Modeling}

Policy-based modeling infers intentions by modeling the strategies or policies that best explain observed actions \cite{zhang2020opponent,yu2022model}. By simulating and refining potential opponent policies, it enables more accurate prediction of opponents' behaviors and goals. Raileanu \textit{et al.} \cite{raileanu2018modeling} propose the self-other modeling scheme, where agents use their own policies to model others and infer the goals of others during online interactions. The approach significantly improves performance in multi-agent tasks by adapting to the behavior of other agents and optimizing based on inferred goals. In adversarial scenarios, it achieves a win rate more than 5 times higher than baseline schemes. Zhang \textit{et al.} \cite{zhang2020opponent} introduce a policy reconstruction method for multi-objective normal form games, where agents infer opponent policies from conditional action frequencies, improving performance by predicting actions more accurately and adjusting strategies for better outcomes in Nash equilibrium.

\subsubsection{Mind Modeling-based Intent Inference}

Mind modeling-based intent inference further focuses on reasoning about the internal mental states that drive agent behavior. Drawing on cognitive concepts such as theory of mind, these approaches infer agents’ beliefs and intentions from observed actions and interactions. This enables a deeper understanding of intentions in complex multi-agent environments. Representative methods include Bayesian mind models, mental state modeling, and LLM-based mind modeling.

\paragraph{Bayesian Mind Modeling}

Bayesian mind modeling infers agents’ intentions by capturing the relationship between their mental states and observable behaviors, leveraging a Bayesian framework for inverse reasoning. Pöppel \textit{et al.} \cite{poppel2018satisficing} introduce a satisficing Bayesian Theory of Mind (ToM) that simplifies complex Bayesian inference by switching between discrete belief states based on the level of surprisal. 
This approach explains agent behavior efficiently, particularly in tasks involving various sources of uncertainty, by balancing computational demands with inferential accuracy. Lim \textit{et al.} \cite{lim2020improving} further propose a Bayesian ToM method for multi-agent cooperation, in which agents model each other’s intentions to support more effective collaborative decision-making. It enhances teamwork by enabling agents to anticipate and adapt to each other’s goals in joint tasks.

\paragraph{Mental State Modeling}

Mental state modeling infers intentions by explicitly representing other agents' mental states, including their beliefs, desires, and intentions \cite{rabinowitz2018machine,shi2025muma}. Rabinowitz \textit{et al.} \cite{rabinowitz2018machine} propose the Theory of Mind neural network (ToMnet), which learns to model agents' mental states from limited behavioral observations. This approach enables the prediction of agent behavior across diverse agent types with minimal data. Wang \textit{et al.} \cite{wang2021tom2c} introduce ToM2C, a multi-agent cooperation method that applies theory of mind to explicitly model the intentions and beliefs of other agents. By relating agents’ mental states to their goals, ToM2C improves collaborative performance. In multi-sensor, multi-target coverage scenarios with restricted local observations, this method achieves a 5\% to 10\% improvement in goal coverage over prior approaches.

\paragraph{LLM-based Mind Modeling}

\begin{figure}
    \centering
    \includegraphics[width=0.8\linewidth]{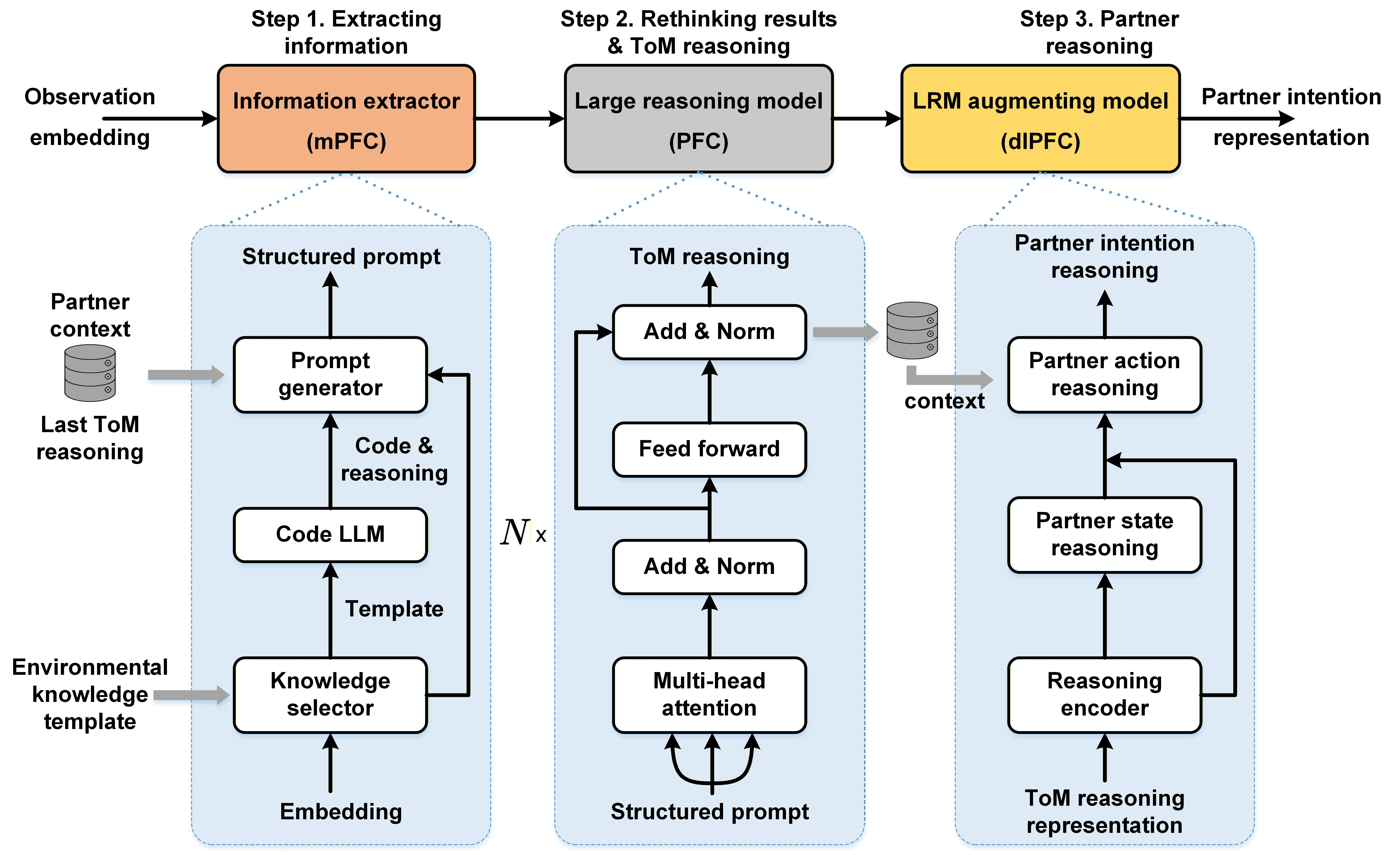}
    \caption{Illustration of the ToM reasoner for partner intention modelling \cite{li2025theory}, consisting of information extraction, ToM reasoning, and partner reasoning stages for constructing structured prompts, generating ToM reasoning, and inferring partner intention representations.}
    \label{fig:ToM modeling}
\end{figure}

LLM-based mind modeling leverages LLMs to infer agents' mental states, exploiting their advanced reasoning capabilities to simulate and predict the intentions of others \cite{li2023theory,cross2024hypothetical}. Cross \textit{et al.} \cite{cross2024hypothetical} introduce Hypothetical Minds, a framework that employs an LLM-based ToM module to hypothesize and evaluate opponents’ strategies in multi-agent settings. This approach yields substantial performance gains in both mixed-motive and collaborative tasks. As illustrated in Figure \ref{fig:ToM modeling}, Li \textit{et al.} \cite{li2025theory} propose an LRM-based ToM framework that infers partners’ beliefs and intentions in cooperative multi-agent tasks. It enhances partner modelling through structured prompting and ToM reasoning, leading to more effective cooperation across diverse scenarios.

\subsection{Semantic Encoding and Processing Layer}
The semantic encoding and processing layer functions as a bridge between agent intention and physical transmission. As illustrated in Table \ref{semantic encoding and processing}, we explore several key enabling technologies integrated within this layer, including semantic-based coding, semantic-based beam management, semantic-based CSI feedback, semantic-based HARQ, AoSI, and semantic KBs.

\begin{table}[!ht]
    \centering
    \caption{Representative approaches for the semantic encoding and processing layer}
    \label{semantic encoding and processing}
	\renewcommand{\arraystretch}{1.5}
	\footnotesize
    \begin{tabular}{>{\arraybackslash}m{0.14\textwidth} >{\arraybackslash}m{0.25\textwidth} >{\arraybackslash}m{0.53\textwidth}}
        \toprule
        \textbf{Category} & \textbf{Sub-Category} & \textbf{Descriptions} \\
        \midrule
        \multirow{2}{=}{Semantic-based Coding}  
        & JSCC & Parameterizes functions of transmitter and receiver using deep neural network to map source data into channel symbols. \cite{bourtsoulatze2019deep, dai2022nonlinear, zhang2023swinjscc, bo2024joint}. \\
        \cmidrule{2-3}
        & Generative Coding & Leverages generative models, such as GAN \cite{lokumarambage2023wireless}, diffusion \cite{xu2025semantic,meng2026secure}, and LLM \cite{salehi2025llm}, to achieve semantic reconstruction based on conditional generation. \\
        \midrule
        \multirow{2}{=}{Semantic-based Beam Management}
        & Visual Semantic-assisted Beam Management & Uses extracted environmental semantics, such as blockage distribution \cite{wen2023vision}, keypoints coordinates \cite{yang2023environment}, and localization \cite{raha2025advancing, khan2025semqnet}, to infer the optimal beam index. \\
        \cmidrule{2-3}
        & Channel Semantic-assisted Beam Management & Combines channel semantics and source semantics to improve beamforming performance \cite{wu2024deep, sun2025towards}. \\
        \midrule
        \multirow{3}{=}{Semantic-based CSI Feedback}
        & Reconstruction-oriented CSI Feedback & Reconstructs CSI by extracting its representative semantic features \cite{xie2024robust, zheng2025semantic, gong2026robust}. \\
        \cmidrule{2-3}
        & Knowledge-driven CSI Feedback & Enhances CSI by leveraging pre-shared prior knowledge, such as semantic label \cite{zhu2024semantic} and channel quality indicator \cite{ren2025semcsinet}, to achieve inference. \\
        \cmidrule{2-3}
        & Optimization for CSI Feedback & Optimizes CSI based on specific task, such as spectral efficiency \cite{gao2023hybrid}, data hiding \cite{cao2023adaptive}, and image reconstruction \cite{zhang2024scan}. \\
        \midrule
        \multirow{3}{=}{Semantic-based HARQ} 
        & Similarity-based HARQ & Enables decision mechanism based on semantic similarity instead of CRC using similarity detection network \cite{jiang2022deep, li2025semantic}. \\
        \cmidrule{2-3}
        & Feature-based HARQ & Identifies retransmitted data based on significant features, such as semantic base\cite{zheng2025semantic2} or importance map\cite{sheng2024semantic}. \\
        \cmidrule{2-3}
        & Adaptive HARQ & Schedules transmission resources adaptively to achieve dynamic mechanism using RL or policy network \cite{zhou2022adaptive}. \\
        \midrule
        \multirow{2}{=}{Age of Semantic Information}
        & Reconstruction-oriented AoSI & Defines AoI in semantic level to minimize the discrepancy between the physical reality and its semantic representation \cite{maatouk2020age, li2024toward, delfani2024semantics}. \\
        \cmidrule{2-3}
        & Task-oriented AoSI & Focuses on the information significance in specific tasks to prioritize intended objective \cite{chiariotti2022query, yates2021age}. \\
        \midrule
        \multirow{2}{=}{Semantic KB}
        & Semantic KB Construction & Defines the structural representation and maintenance mechanisms to construct semantic KB \cite{wang2024unified, ren2024knowledge}. \\
        \cmidrule{2-3}
        & Semantic KB Deployment & Leverages shared KBs to empower transmission tasks, such as indexing \cite{EndToEndGenerativeSKB}, residual compression \cite{DeepLearningEmpoweredSharedKB}, GraphRAG-assisted subgraph extraction\cite{fan2025kgrag}, and logical error correction based on triples\cite{CognitiveSemComKG}. \\
        \bottomrule      
    \end{tabular}
\end{table}

\subsubsection{Semantic-based Coding}
\begin{figure}
    \centering
    \includegraphics[width=0.8\linewidth]{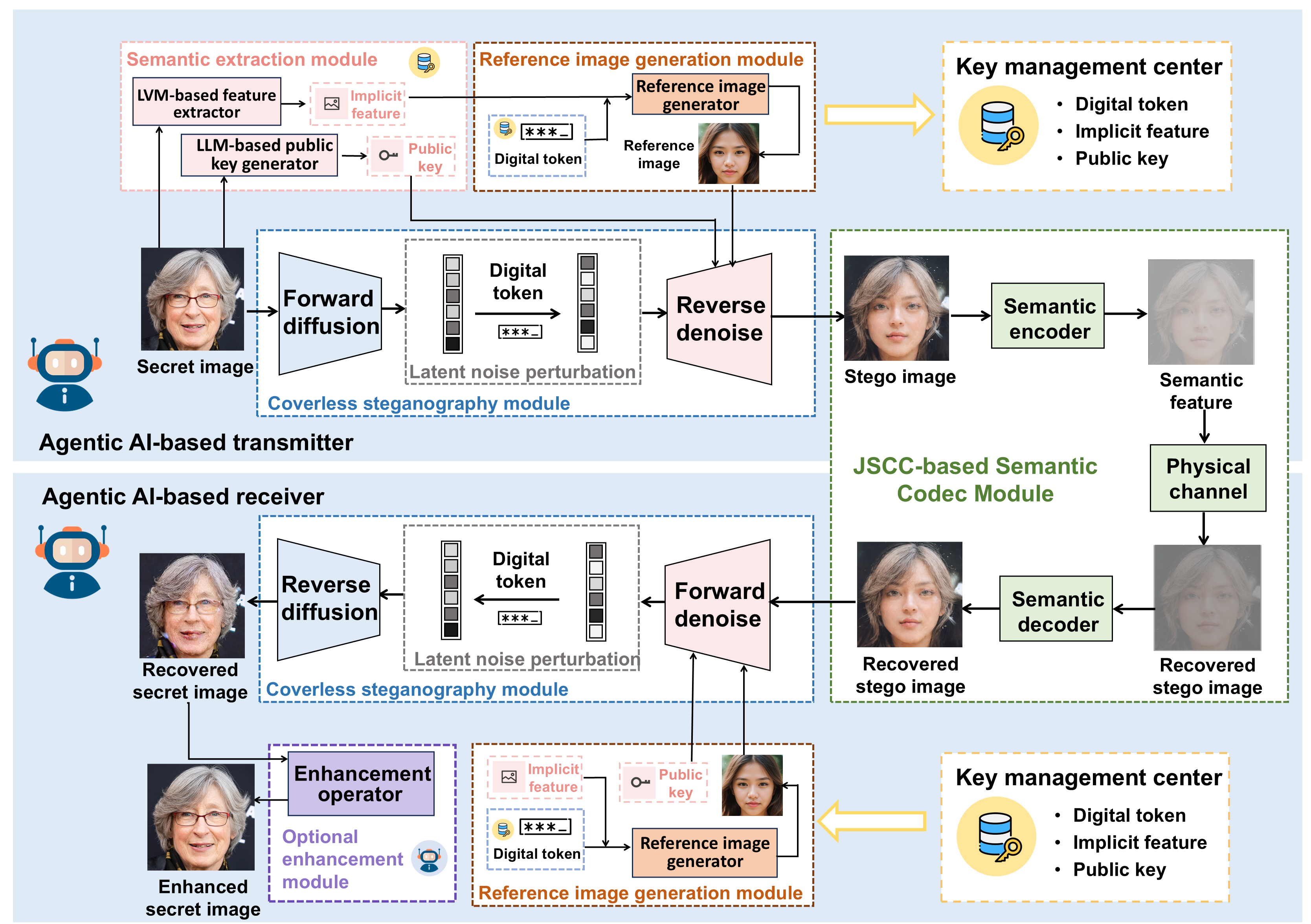}
    \caption{Illustration of agentic AI-driven semantic steganography communication \cite{meng2026secure}, which includes semantic extraction, digital token controlled reference image generation, coverless steganography, semantic codec, and optional task-oriented enhancement modules.}
    \label{fig:coding}
\end{figure}

To address the cliff effect inherent in conventional separation-based coding schemes, where slight fluctuations in channel quality can lead to a catastrophic breakdown in decoding reliability, semantic-based coding adopts an end-to-end learnable strategy integrating feature extraction, source compression, and channel coding. This paradigm shifts the focus from bit-level accuracy to the preservation of core semantic meaning, significantly enhancing transmission robustness in adverse environments. Semantic-based coding schemes are generally divided into two primary categories: Joint Source-Channel Coding (JSCC) and generative coding.
\paragraph{JSCC}
JSCC utilizes neural networks to directly map multi-modal source data into channel symbols to build the connection between transmitter and receiver. By jointly optimizing the transmitter and receiver, JSCC achieves effective performance gains even in low SNR environments. Bourtsoulatze \textit{et al.} \cite{bourtsoulatze2019deep} first propose a deep JSCC technique for wireless image transmission that eliminates the need for separate compression or error correction coding. By parameterizing the encoder and decoder functions with convolutional neural networks (CNNs), the scheme directly maps image pixel values to complex-valued channel input symbols.  Furthermore, Yang \textit{et al.}\cite{zhang2023swinjscc} propose SwinJSCC, a novel neural JSCC backbone that integrates the Swin Transformer to overcome the limited capabilities of traditional CNN-based models. They introduce channel ModNet and rate ModNet to scale latent representations based on CSI and target transmission rates, achieving superior performance and faster end-to-end coding speeds. Additionally, Bo \textit{et al.}\cite{bo2024joint} propose a joint coding-modulation scheme that maps source data to discrete constellation points, enabling digital SemCom.
\paragraph{Generative Coding}
Generative coding leverages the prior knowledge embedded in advanced generative models to reconstruct transmitted information by framing the semantic decoding as a conditional generation task. Xu \textit{et al.} \cite{xu2025semantic} propose a latent diffusion model-based scheme, which employs a joint semantic equalizer and denoiser module to recover clean semantic features and mitigate channel effects. As depicted in Figure \ref{fig:coding}, Meng \textit{et al.} \cite{meng2026secure} propose an agentic AI-driven semantic steganography communication scheme that hides a secret image into a stego image based on a diffusion model to achieve invisible encryption, which realizes secure SemCom through generative coding. Meanwhile, Salehi \textit{et al.} \cite{salehi2025llm} introduce the KG-LLM framework, which integrates KG extraction for structured compression with LLM coding for contextualized semantic representation.
\subsubsection{Semantic-based Beam Management}
Semantic-based beam management aims to optimize beamforming by incorporating semantic information derived from the environment or the source data, thereby enhancing efficiency and robustness to ensure stable agent communication even in highly dynamic scenarios. It can be divided into visual semantic-assisted and channel semantic-assisted beam management.
\paragraph{Visual Semantic-assisted Beam Management}
Visual semantic-assisted beam management relies on high-level environmental cues, such as blockage distribution, keypoint coordinates, and localization, to proactively predict and maintain optimal beam alignment. Yang \textit{et al.} \cite{yang2023environment} extract semantic information from environmental images captured by street cameras and selectively encode it based on task relevance to support channel-related decision-making. By predicting the optimal beam index and potential blockage states without pilot training or costly beam sweeps, the proposed scheme enables efficient beam management. Furthermore, Wen \textit{et al.} \cite{wen2023vision} define environmental semantics as the spatial distribution of scatterers that influence the wireless channel and employ keypoint detection techniques to extract these features from raw images, subsequently mapping them to optimal beam pairs.
\paragraph{Channel Semantic-assisted Beam Management}
Channel semantic-assisted beam management focuses on the semantics of the source content to ensure that beamforming vectors are optimized based on the significance and characteristics of the transmitted information. By integrating source importance into the spatial domain, it achieves superior transmission efficiency and task-oriented performance. Wu \textit{et al.} \cite{wu2024deep} propose a deep joint semantic coding and beamforming scheme. Specifically, the authors utilize two specialized semantic extraction networks to extract features from both the image source and CSI, and introduce hybrid data-driven and model-driven semantic-aware beamforming networks that jointly optimize coding and spatial processing. 

\subsubsection{Semantic-based Channel State Information (CSI) Feedback}
CSI is essential for agents to enable channel-adaptive transmission. However, the massive dimensionality of channels imposes excessive feedback and computational overhead. Semantic-based CSI feedback addresses these challenges by extracting and transmitting only task-relevant channel features, ensuring robust and goal-oriented performance across dynamic wireless environments. Representative approaches include reconstruction-oriented CSI feedback, knowledge-driven CSI feedback, and optimization for CSI feedback.

\paragraph{Reconstruction-oriented CSI Feedback}
A line of research focuses on reconstructing the CSI matrix by  extracting its representative semantic features. Xie \textit{et al.}  \cite{xie2024robust} propose a learnable CSI fusion SemCom framework that employs an attention masking map and treats MIMO CSI as side information to enhance robustness. Similarly, Gong \textit{et al.} \cite{gong2026robust} introduce a channel matrix adaptor that operates alongside the channel codec to compensate for misaligned CSI, thereby mitigating reconstructed errors between the estimated and actual channel matrices.

\paragraph{Knowledge-driven CSI Feedback}
Knowledge-driven CSI feedback leverages prior knowledge shared between the transmitter and receiver to achieve accurate channel inference. Ren \textit{et al.} \cite{ren2025semcsinet} propose SemCSINet, a semantic-aware Transformer-based framework that incorporates the channel quality indicator as a semantic prior to guide the feedback loop in massive MIMO systems, significantly improving reconstruction accuracy and system robustness. Meanwhile, Zhu \textit{et al.} \cite{zhu2024semantic} employ clustering methods to predefine a semantic label database for CSI feedback. It maps high-dimensional CSI into concise semantic labels and transmits only the label corresponding to the current CSI, substantially reducing feedback overhead while preserving task accuracy.

\paragraph{Optimization for CSI Feedback}
Rather than pursuing full channel reconstruction, this approach optimizes the feedback process based on specific objectives and extracts only the channel features critical to the current task. Cao \textit{et al.} \cite{cao2023adaptive} introduce an adaptive CSI feedback framework based on the information bottleneck principle, optimized for multi-task efficiency and data privacy. A key innovation lies in the hidden transfer of sensory data within the CSI feedback loop, thereby eliminating the additional resource overhead typically required for separate data reporting.

\subsubsection{Semantic-based Hybrid Automatic Repeat reQuest (HARQ)}
\begin{figure}
    \centering
    \includegraphics[width=0.5\linewidth]{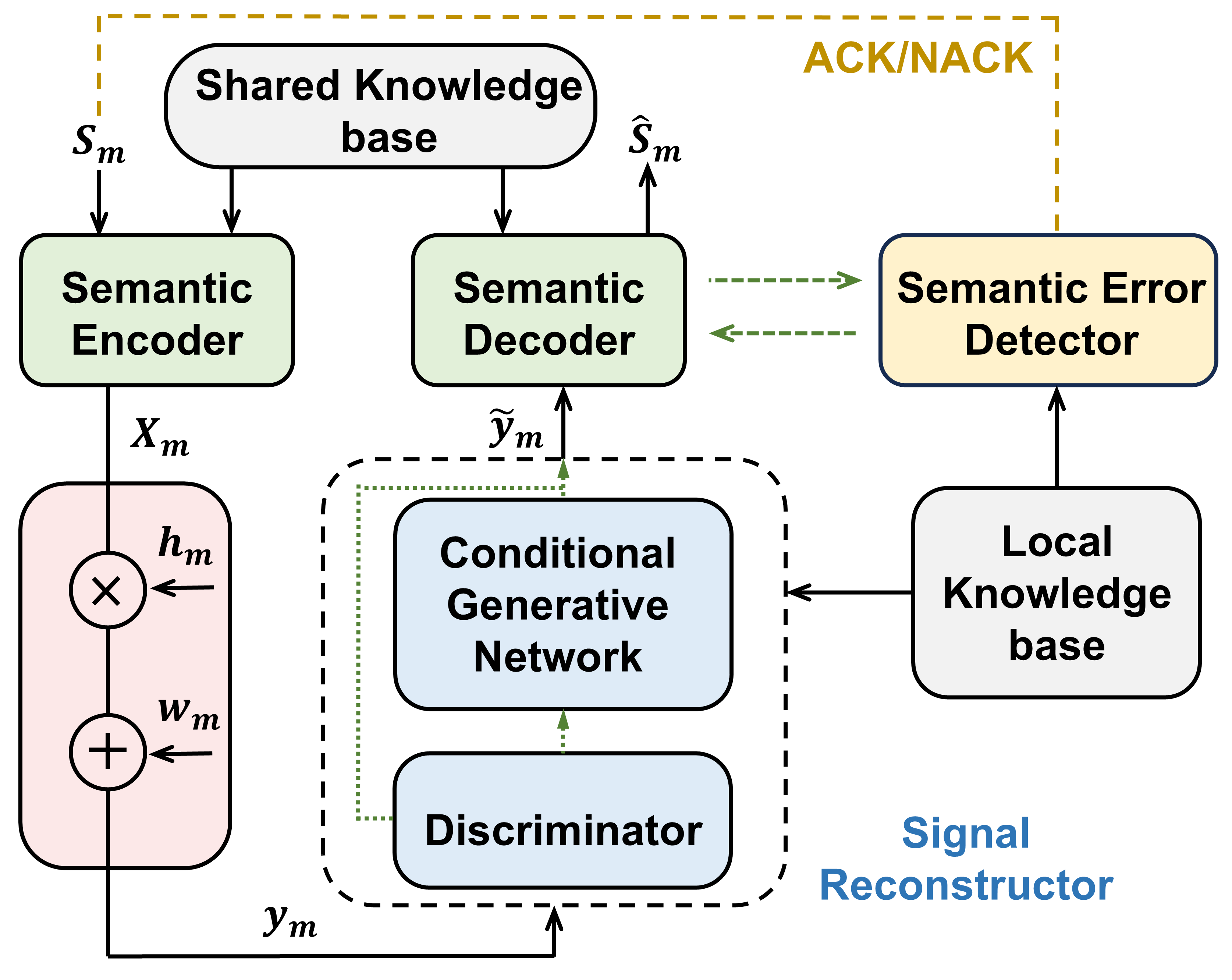}
    \caption{Illustration of the generative semantic HARQ framework \cite{li2025semantic}, which includes the knowledge base, the semantic communication system, and the HARQ enhancement module}
    \label{fig:harq}
\end{figure}
Semantic-based HARQ shifts the reliability paradigm from bit-level error correction to the preservation of meaning. By integrating semantic awareness into feedback and retransmission processes, it optimizes transmission efficiency and ensures reconstruction quality. It can be categorized into similarity-based HARQ, feature-based HARQ, and adaptive HARQ.
\paragraph{Similarity-based HARQ}
Similarity-based HARQ replaces bit-level checks with semantic similarity metrics to determine whether retransmission is needed, thereby avoiding unnecessary overhead. Jiang \textit{et al.} \cite{jiang2022deep} propose a Sim32 framework that employs a similarity detection network in place of the conventional cyclic redundancy check. This approach allows the receiver to accept packets that contain bit errors but remain semantically accurate, significantly reducing the number of required retransmissions and conserving communication resources in low-SNR regimes. Additionally, Li \textit{et al.} \cite{li2025semantic} propose a generative semantic HARQ framework tailored for intelligent transportation systems, as shown in Figure \ref{fig:harq}. It introduces a synonymous combining strategy that leverages semantic distance and local knowledge bases, enabling the receiver to autonomously recover information by identifying semantically equivalent content in challenging vehicular environments.
\paragraph{Feature-based HARQ}
Feature-based HARQ enhances transmission reliability by identifying and retransmitting specific semantic components or significant features. Zheng \textit{et al.} \cite{zheng2025semantic2} develop a HARQ mechanism based on a semantic base, where the system precisely identifies and retransmits only the erroneous semantic elements using contextual correlations. Meanwhile, Sheng \textit{et al.} \cite{sheng2024semantic} introduce an importance map-guided HARQ for cooperative perception, which extracts critical semantic information to ensure the reliable transmission of task-essential features, achieving robust perception performance with minimal data overhead.
\paragraph{Adaptive HARQ}
Adaptive HARQ focuses on the intelligent scheduling of transmission resources to dynamically balance reliability and efficiency. Zhou \textit{et al.} \cite{zhou2022adaptive} propose an adaptive bit rate control scheme for incremental knowledge-based HARQ, which employs a policy network trained via RL. By sensing real-time channel conditions and content complexity, the framework adaptively determines the optimal initial transmission length and incremental redundancy steps, effectively optimizing the trade-off between reconstruction distortion and transmission latency.
\subsubsection{Age of Semantic Information}
Driven by the demand for real-time intelligence, the Age of Information (AoI) has emerged as a fundamental metric for quantifying information freshness, typically relying on fixed or periodic update policies to maximize throughput or minimize end-to-end delay. The AoSI extends the traditional AoI by incorporating information significance and task effectiveness into the freshness metric. It shifts the focus from purely temporal freshness to whether the information remains semantically valuable and relevant to the receiver's objectives. It mainly includes reconstruction-oriented and task-oriented AoSI.
\paragraph{Reconstruction-oriented AoSI}
Reconstruction-oriented AoSI aims to minimize the semantic discrepancy between the source and the monitor, ensuring the receiver maintains a high-fidelity representation of the physical state. Rather than measuring elapsed time, it quantifies the duration during which the receiver’s knowledge is inconsistent with the actual state. Maatouk \textit{et al.} \cite{maatouk2020age} propose the Age of Incorrect Information (AoII), which alerts the system only when the monitor holds incorrect or outdated information, effectively shifting the focus from fresh updates to fresh ``informative" updates. Furthermore, Li \textit{et al.} \cite{li2024toward} introduce a goal-oriented tensor framework to integrate various semantic metrics such as AoII and the Value of Information (VoI). Additionally, Delfani \textit{et al.} \cite{delfani2024semantics} explore Query Version AoI (QVAoI) in energy-harvesting systems, optimizing transmission policies to deliver the most significant updates while maintaining energy sustainability.

\paragraph{Task-oriented AoSI}
Task-oriented AoSI focuses on the information relevance to specific downstream tasks. It posits that information freshness is critical only at the moments of task execution, thereby prioritizing updates that directly serve the intended goal. Chiariotti \textit{et al.} \cite{chiariotti2022query} propose the Query AoI (QAoI) for pull-based communication scenarios where information is consumed only upon query generation. By optimizing QAoI, the authors significantly reduce the perceived age at relevant instants for both periodic and stochastic queries, demonstrating superior resource efficiency compared to traditional AoI-based scheduling. Moreover, Yates \cite{yates2021age} introduces the concept of version age in gossip networks, which measures how many versions out of date a node's knowledge is relative to the source. This discrete metric effectively characterizes the timeliness of information diffusion in distributed networks, ensuring that nodes prioritize the most recent versions to achieve collective task objectives.

\subsubsection{Semantic Knowledge Base}
\begin{figure}
    \centering
    \includegraphics[width=0.8\linewidth]{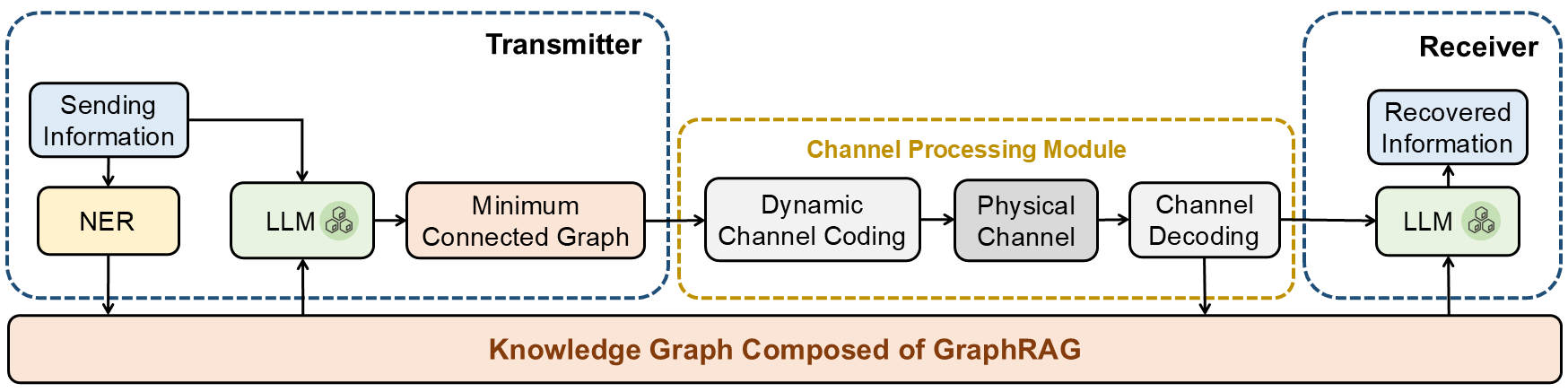}
    \caption{Illustration of the knowledge
    graph-assisted SemCom framework \cite{fan2025kgrag}, which shows the semantic extraction at the transmitter and text reconstruction at the receiver.}
    \label{fig:database}
\end{figure}
The semantic KB stores semantic logic and relationships, enabling effective semantic representation through integrated processing, memory, and reasoning capabilities. Research in this area can be categorized into the construction and deployment of semantic KBs.
\paragraph{Semantic KB Construction}
Semantic KB construction focuses on structural representations and efficient maintenance mechanisms to model complex data attributes and task requirements. Ren \textit{et al.}\cite{ren2024knowledge} propose a generative semantic KB architecture that partitions the knowledge space into three specialized sub-KBs: source, task, and channel KBs. By introducing semantic metalets to standardize units such as embedding vectors, this scheme parameterizes source messages into low-dimensional spaces, effectively bridging the gap between raw data and semantic meaning. Similarly, Wang \textit{et al.} \cite{wang2024unified} develop a unified hierarchical semantic KB framework designed for multi-task scenarios. It employs horizontal construction to maximize the semantic representation space and vertical construction to exploit cross-task correlations through a deep K-subspace clustering method, achieving a significant improvement in knowledge search efficiency for complex reconstruction tasks.
\paragraph{Semantic KB Deployment}
Semantic KB deployment investigates the practical effectiveness of shared KBs in empowering specific transmission tasks, with an emphasis on indexing, compression, and enhanced reasoning capabilities.
To optimize transmission efficiency, Yan \textit{et al.}\cite{EndToEndGenerativeSKB} employ shared KBs to implement a generative encoding-decoding paradigm, where transmitted indices and minimal residual data act as ``prompts" to trigger high-fidelity content synthesis at the receiver. For robust semantic delivery, Hu \textit{et al.} \cite{DeepLearningEmpoweredSharedKB} introduce VQ-VAE-enabled codebooks to extract intrinsic features, reducing the statistical discrepancy between source messages and training examples and thereby mitigating the impact of semantic noise. Meanwhile, to further refine reasoning capabilities, Fan \textit{et al.} \cite{fan2025kgrag} propose a GraphRAG-assisted subgraph extraction method, as illustrated in Figure \ref{fig:database}. It identifies the minimum connected subgraph within a KG to provide precise context for semantic interpretation, significantly reducing bandwidth consumption. Additionally, Zhou \textit{et al.} \cite{CognitiveSemComKG} apply KGs to logical error correction, ensuring consistency of the transmitted information.

\subsection{Distributed Autonomy and Collaboration Layer}
\begin{table*}[!ht]
    \centering
    \caption{Representative approaches for the distributed autonomy and collaboration layer}
    \label{Distributed_Collaboration}
    \renewcommand{\arraystretch}{1.5}
    \footnotesize
    \begin{tabular}{>{\raggedright\arraybackslash}p{0.16\textwidth} >{\raggedright\arraybackslash}p{0.22\textwidth} >{\raggedright\arraybackslash}p{0.54\textwidth}}
        \toprule
        \textbf{Category} & \textbf{Sub-Category} & \textbf{Descriptions} \\
        \midrule
        Distributed Access 
        & MDMA & Allocates unique semantic encoder-decoder pairs or leverages orthogonal embeddings to decouple multi-agent access in the semantic space \cite{zhang2023model, liang2024orthogonal, cao2026s}. \\
        \cmidrule{2-3}
        & Semantic Fusion & Fuses multi-modal or multi-user semantic representations at the access level to resolve ambiguity and reduce transmission redundancy \cite{zhu2024multi, li2022crossmodal, tong2024multi}. \\
        \midrule
        Knowledge Collaboration 
        & Federated Semantic Learning & Enables agents to collaboratively train and update semantic codec models in a resource-efficient and personalized manner without sharing raw data \cite{li2025decentralized, liu2023efficient, peng2024personalized}. \\
        \cmidrule{2-3}
        & Multi-Agent Alignment & Facilitates semantic interoperability and background KB consensus among diverse autonomous agents via signaling games or curriculum learning \cite{choi2022semantic, farshbafan2023curriculum, rosic2025semantic}. \\
        \cmidrule{2-3}
        & Semantic Relaying & Employs intermediate nodes for progressive semantic feature computation, denoising, and forwarding over long distances \cite{zhao2024resource, hu2023multiuser, liu2023semantic}. \\
        \cmidrule{2-3}
        & Collaborative Inference & Partitions semantic extraction and task execution workloads among edge devices and servers for single-task or multi-task joint knowledge inference \cite{lo2023collaborative, shao2022task, razlighi2025cooperative}. \\
        \midrule
        Resource Scheduling
        & Query-Semantic Scheduling & Optimizes communication and computation resources jointly, driven top-down by specific downstream tasks or user query intents \cite{cai2025query, yan2024qoe, zeng2023task}. \\
        \cmidrule{2-3}
        & Semantic Importance-Aware Scheduling & Prioritizes the allocation of physical resources bottom-up based on the spatial-temporal importance of semantic features \cite{wang2024feature, chen2024multi, wang2022performance}. \\
        \cmidrule{2-3}
        & Adaptive Semantic Scheduling & Dynamically controls the semantic compression ratio and hybrid transmission modes in response to real-time channel fluctuations \cite{liu2023adaptable, zhu2022semantic, xia2024wireless}. \\
        \bottomrule
    \end{tabular}
\end{table*}

To support massive connectivity and autonomous interaction of AI agents in future 6G networks, the traditional bit-level centralized communication architecture must evolve into a semantic-based distributed paradigm. As summarized in Table \ref{Distributed_Collaboration}, we thoroughly investigate representative approaches for the distributed autonomy and collaboration layer, which includes distributed access, knowledge collaboration, and resource scheduling.

\subsubsection{Distributed Access}

By shifting from bit-level orthogonal multiplexing to semantic-level access, distributed access ensures the efficient utilization of limited spectrum resources among massive agents. It mainly includes MDMA and semantic fusion.

\begin{figure}
    \centering
    \includegraphics[width=0.8\linewidth]{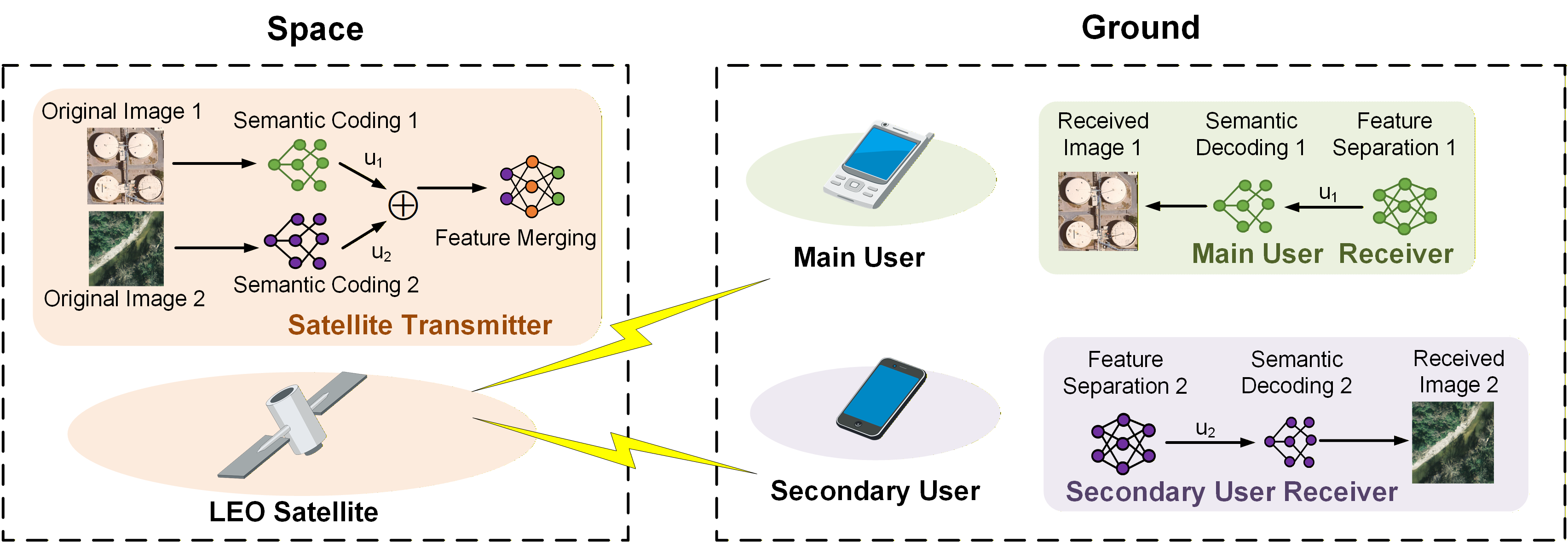}
    \caption{System model of multiuser satellite-ground SemCom\cite{cao2026s}, illustrating the collaborative framework among the LEO satellite, semantic transmission links, and diverse ground terminals for efficient feature-based communication.}
    \label{fig:S-MDMA}
\end{figure}

\paragraph{Model Division Multiple Access}
Traditional multiple access techniques partition physical resources across time, frequency, or code domains, often leading to severe capacity bottlenecks and a pronounced ``cliff effect'' in dense agent communication networks. MDMA transforms this paradigm by multiplexing users directly in a high-dimensional semantic feature space. Zhang \textit{et al.} \cite{zhang2023model} first establish the foundational MDMA framework, which allocates unique semantic encoder-decoder pairs for each user. This architecture is grounded in the assumption that different AI models possess natural separability, enabling a BS to distinguish superimposed signals by decoding them through user-specific models without requiring extra bandwidth.
However, as the number of agents grows, overlapping semantic features can cause significant mutual interference. To address this, subsequent advances in MDMA aim to achieve strictly interference-free concurrency. For example, Orthogonal-MDMA (O-MDMA) leverages the inherent noise resilience of semantic models to suppress multi-user interference through structured subspace projections \cite{liang2024orthogonal}. Furthermore, addressing the stringent bandwidth constraints and highly dynamic topologies in satellite-ground links, Cao \textit{et al.} \cite{cao2026s} propose Sensitivity-aware MDMA (S-MDMA), which is shown in Figure \ref{fig:S-MDMA}. This approach extracts and merges shared semantic features and applies a sensitivity-based sorting algorithm to retain only the most critical structural components. By mapping shared and unique features into mutually orthogonal subspaces via Kronecker-based embedding, S-MDMA eliminates inter-user interference entirely.
Simulation results demonstrate that it maintains a Peak Signal-to-Noise Ratio (PSNR) above 28 dB and a structural similarity index (SSIM) exceeding 0.95 even under severely degraded conditions.

\paragraph{Semantic Fusion}
Semantic fusion functions as a key access-level mechanism that synthesizes redundant or complementary data generated by multiple agents. By merging data before transmission, it resolves modal ambiguities and substantially reduces communication overhead over the air interface. Zhu \textit{et al.} \cite{zhu2024multi} develop a multi-modal fusion framework that integrates diverse sensory inputs, such as visual, auditory, and textual data, directly at the edge device. This integration enables the agent to transmit a unified, low-dimensional semantic representation capable of supporting multiple downstream tasks simultaneously. To further reduce ambiguity across modalities, Li \textit{et al.} \cite{li2022crossmodal} introduce highly reliable cross-modal mapping mechanisms that ensure consistent semantic extraction by projecting heterogeneous data types into a shared latent space.
In multi-agent scenarios, physical-layer fusion strategies can merge overlapping semantic features among neighboring agents. For example, Tong \textit{et al.} \cite{tong2024multi} propose a multi-user semantic fusion strategy tailored for degraded broadcast channels. By performing feature-level aggregation directly at the access point, this approach exploits the superposition property of wireless channels to combine highly correlated semantic elements, thereby eliminating redundant background information. Simulation results indicate that such collaborative fusion schemes can reduce the overall transmission overhead by up to 40\% while preserving the semantic fidelity required for downstream perception tasks.

\subsubsection{Knowledge Collaboration}
Knowledge collaboration establishes a closed-loop lifecycle encompassing model consensus building, context alignment, physical-layer transmission, and joint execution. It enables isolated agents to form a cohesive, distributed intelligence network.

\begin{figure}
    \centering
    \includegraphics[width=0.8\linewidth]{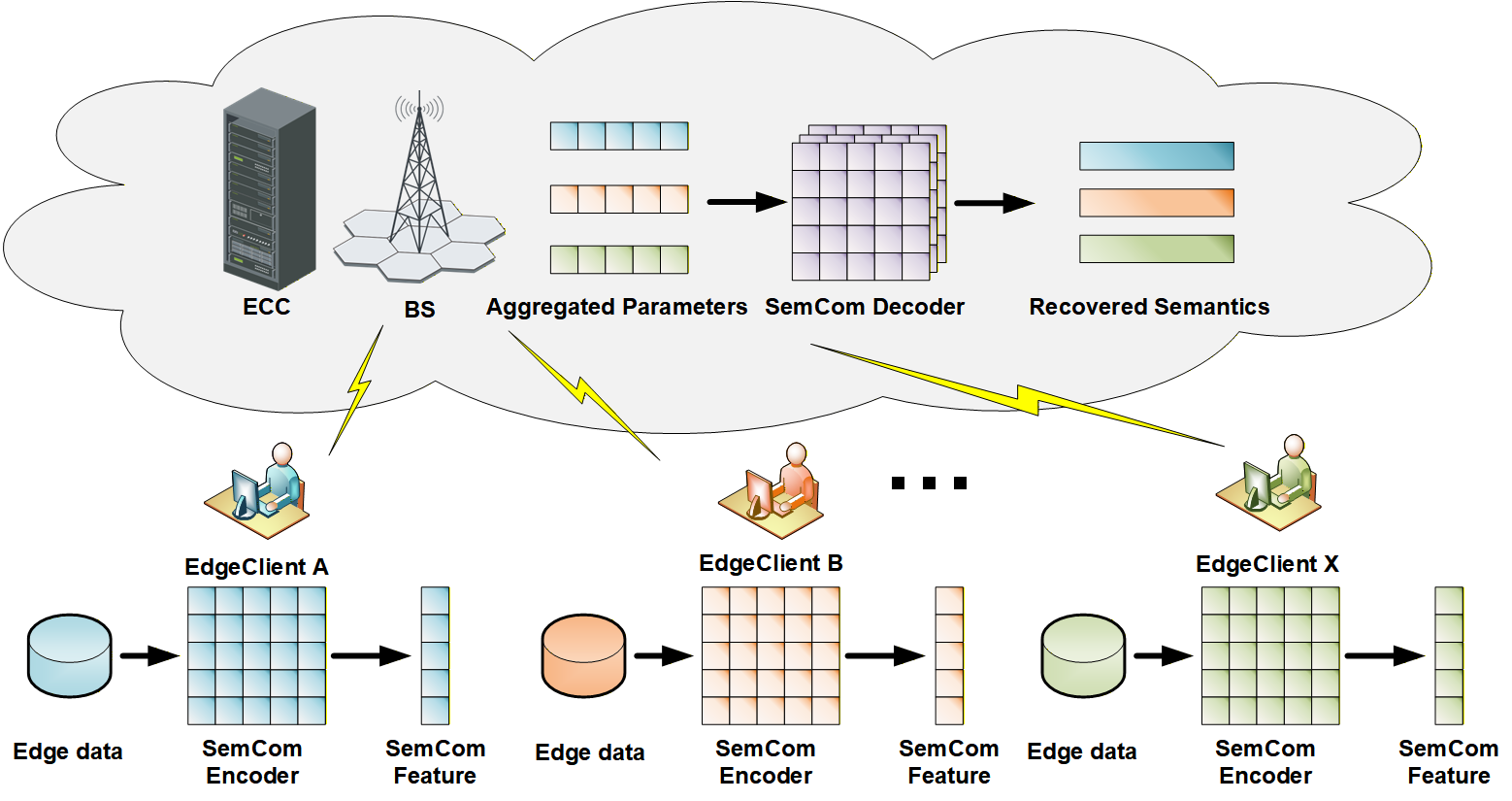}
    \caption{The decentralized semantic federated learning framework \cite{li2025decentralized}, illustrating the integration of local semantic encoding at edge clients and centralized parameter aggregation at the base station for efficient data recovery.}
    \label{fig:Knowledge Collaboration - FSL}
\end{figure}

\paragraph{Federated Semantic Learning}
Federated semantic learning facilitates the collaborative training and updating of semantic codec models across agents without sharing raw privacy-sensitive data, forming the foundation of distributed semantic-level intelligence. By exchanging only model gradients or weights, agents can collectively build a generalized semantic representation space. As shown in Figure \ref{fig:Knowledge Collaboration - FSL}, Li \textit{et al.} \cite{li2025decentralized} propose a decentralized semantic federated learning architecture tailored for real-time public safety tasks, effectively avoiding the single point of failure and data centralization bottlenecks inherent in traditional cloud-based training.
To mitigate the significant communication overhead associated with transmitting large neural network parameters over wireless channels, Liu \textit{et al.} \cite{liu2023efficient} introduce a resource-aware allocation and semantic extraction scheme specifically designed for federated semantic learning-empowered vehicular networks. Furthermore, addressing the highly heterogeneous data distributions (Non-IID) across different physical environments, Peng \textit{et al.} \cite{peng2024personalized} develop a personalized federated learning framework. It integrates global knowledge aggregation with local micro-model fine-tuning, enabling agents to maintain customized local models that adapt effectively to specific environmental changes while still benefiting from collaborative intelligence.

\paragraph{Multi-Agent Alignment}
Multi-agent alignment resolves discrepancies arising from disparate background KBs, serving as an essential prerequisite for accurate semantic encoding and decoding. Theoretical frameworks model this alignment process as signaling games, where agents with correlated KBs negotiate and converge on a common semantic language through continuous mathematical interaction \cite{choi2022semantic}. For more complex environments, Farshbafan \textit{et al.} \cite{farshbafan2023curriculum} employ reinforcement-based curriculum learning to gradually adapt the transmission strategies of goal-oriented agents, actively minimizing the semantic ambiguity caused by mismatched local contexts.
In practical implementations, these mechanisms are realized through explicit KB synchronization protocols to ensure semantic interoperability. For instance, Rosic \textit{et al.} \cite{rosic2025semantic} address semantic interoperability in autonomous maritime domains, proposing robust mechanisms to dynamically align the KGs of disparate marine agents. By aligning these distributed knowledge structures prior to execution, simulation results demonstrate that agents can achieve over a 30\% improvement in semantic recovery accuracy and significantly reduce operational conflicts when navigating in previously unseen collaborative environments.

\begin{figure}
    \centering
    \includegraphics[width=0.8\linewidth]{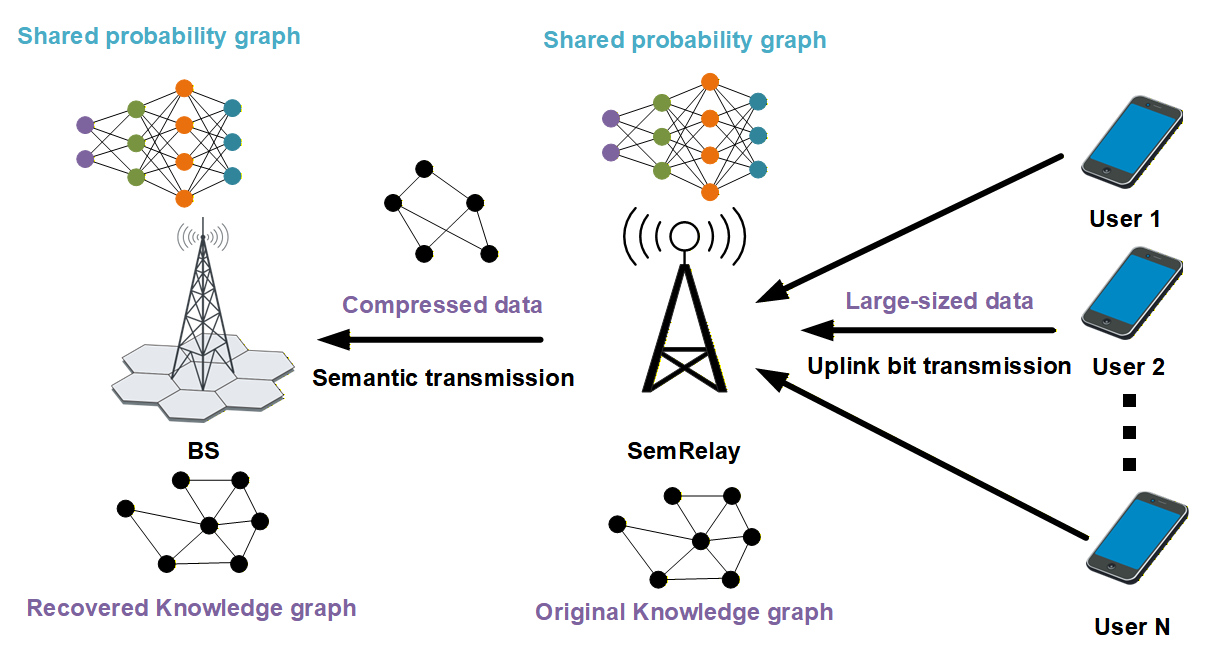}
    \caption{The SemRelay-aided semantic communication system\cite{zhao2024resource}, illustrating the uplink bit transmission from users to the SemRelay and the subsequent semantic transmission to the base station based on shared probability graphs.}
    \label{fig:Knowledge Collaboration - Relaying}
\end{figure}

\paragraph{Semantic Relaying}
Semantic relaying transforms intermediate nodes into intelligent entities for progressive feature computation, denoising, and regeneration, thereby extending the coverage of knowledge exchange. Rather than simply amplifying and forwarding noisy analog signals, these relays actively decode, refine, and re-encode the semantic features to combat channel fading. As illustrated in Figure \ref{fig:Knowledge Collaboration - Relaying}, Zhao \textit{et al.} \cite{zhao2024resource} utilize probabilistic graphs to mathematically model state transitions at semantic relays, proposing a joint communication and computation resource allocation scheme that maximizes cooperative reliability over multiple hops.
Addressing the complexities of multi-user contention, Hu \textit{et al.} \cite{hu2023multiuser} formulate a non-convex optimization problem to optimally divide bandwidth and computational resources among multiple users relying on a shared semantic relay. Additionally, Liu \textit{et al.} \cite{liu2023semantic} expand the optimization scope by jointly designing the physical placement of semantic relays and the associated bandwidth allocation.

\paragraph{Collaborative Inference}
Collaborative inference partitions heavy semantic extraction and deep learning workloads among resource-constrained agents and powerful edge servers, transforming isolated processing into distributed joint reasoning. The paradigm has evolved from simple one-to-one splitting to complex multi-view aggregation. Lo \textit{et al.} \cite{lo2023collaborative} initially investigate a device-edge collaboration scheme in which the agent executes the early layers of a neural network to extract lightweight semantic features, offloading computationally intensive classification layers to the edge server. Extending this to multi-agent settings, Shao \textit{et al.} \cite{shao2022task} propose a task-oriented communication framework for multi-device cooperative edge inference. Here, the edge server aggregates spatially diverse, multi-view semantic features from distributed cameras or sensors to enhance the global detection accuracy.
Recent advances have led to frameworks that enable clusters of agents to extract generalized features for simultaneous, heterogeneous reasoning. Zhu \textit{et al.} \cite{razlighi2025cooperative} introduce a cooperative and collaborative multi-task SemCom framework for distributed sources. It allows a cluster of agents to extract generalized semantic features cooperatively, simultaneously supporting diverse downstream reasoning tasks such as object detection and semantic segmentation, highlighting the efficiency of distributed multi-task execution.

\subsubsection{Resource Orchestration and Scheduling}
With access and collaboration mechanisms established, the network must optimally allocate its limited physical resources. In semantic-based agent communication networks, resource scheduling shifts fundamentally from maximizing bit-rate to maximizing the effective delivery of meaning. This process follows the logic: defining the intent, evaluating content importance, and adapting to the physical channel conditions. Representative methods include query-semantic scheduling, semantic importance-aware scheduling, and adaptive semantic scheduling.

\paragraph{Query-Semantic Scheduling}
Query-semantic scheduling allocates computing and communication resources based explicitly on the diverse requirements and priorities of the downstream tasks requested by agents. Cai \textit{et al.} \cite{cai2025query} propose a query-aware semantic encoder-based resource allocation policy that dynamically schedules network slices by precisely matching the extracted data stream characteristics to specific user querying intents. 
Similarly, focusing on user-centric requirements, Yan \textit{et al.} \cite{yan2024qoe} formulate a quality of experience (QoE)-based resource allocation algorithm for multi-task networks. It optimizes power and channel assignments to maximize subjective task satisfaction rather than objective bit-rates. For mission-critical scenarios, Zeng \textit{et al.} \cite{zeng2023task} develop a task-oriented SemCom scheme utilizing rate-splitting multiple access. Instead of treating all data equally, this approach focuses strictly on the effectiveness level of communication. Simulation results demonstrate that it can increase overall QoE scores by over 20\% in real-time control applications while satisfying ultra-reliable and low-latency communication constraints.

\paragraph{Semantic Importance-Aware Scheduling}
Once task intent is defined, the scheduling logic transitions to a content-driven phase. Semantic importance-aware scheduling prioritizes physical resource allocation by evaluating the spatial-temporal value of extracted semantic features, ensuring that bandwidth and power are directed primarily toward critical information. To quantify this value, Wang \textit{et al.} \cite{wang2024feature} introduce a feature importance-aware semantic transmission strategy that dynamically allocates power blocks by calculating the precise contribution of each semantic symbol to final classification accuracy using attention mechanisms. 
From a temporal perspective, Chen and Gong \cite{chen2024multi} utilize the AoSI metric to design a multi-source scheduling algorithm, prioritizing the transmission of fresh and highly impactful semantic updates to prevent knowledge staleness. Furthermore, Wang \textit{et al.} \cite{wang2022performance} employ an attention-based deep RL approach to automate this evaluation, enabling the BS to dynamically learn the hidden correlations between semantic importance distributions and channel conditions. 


\paragraph{Adaptive Semantic Scheduling}
scheduling strategies must contend with highly dynamic physical environments. Adaptive semantic scheduling operates at the execution level, dynamically adjusting semantic compression ratios and switching transmission modes in response to real-time CSI fluctuations. Liu \textit{et al.} \cite{liu2023adaptable} propose an adaptable semantic compression algorithm that optimally balances local computational latency against transmission delay by adjusting the dimension of semantic vectors based on real-time SNR. 
For high-throughput applications, Zhu \textit{et al.} \cite{zhu2022semantic} design an adaptive control scheme for volumetric video services, dynamically altering compression ratios to maintain visual QoE amidst severe bandwidth constraints. Moreover, recognizing that pure SemCom may not be optimal in high SNR environments, Xia \textit{et al.} \cite{xia2024wireless} propose a hybrid bit/semantic communication optimization framework. This scheme allows the network to dynamically switch agents between traditional bit-level transmission and semantic extraction modes depending on current channel quality. Such real-time adaptation ensures system resilience, providing robust and highly adaptable scheduling closure for the entire agent communication network even during severe channel fading events.

\section{Semantic-enhanced Agentic AI Systems}
\label{sectionagenticai}

This section summarizes how semantics enhance the four stages of agentic AI systems: perception, memory, reasoning, and action.

\subsection{Semantic-based Perception Stage}
\begin{table}
    \centering
    \caption{Representative approaches for the semantic-based perception stage}
    \label{Perception_Stage}
	\renewcommand{\arraystretch}{1.5}
	\footnotesize
    \begin{tabular}{>{\arraybackslash}m{0.14\textwidth} >{\arraybackslash}m{0.23\textwidth} >{\arraybackslash}m{0.55\textwidth}}
        \toprule
        \textbf{Category} & \textbf{Sub-Category} & \textbf{Descriptions} \\
        \midrule
        \multirow{2}{=}{Semantic Feature Extraction and Representation}  
        & Unified Multi-modal Embedding & Maps heterogeneous data into a shared semantic space to enable emergent alignment and cross-modal correlation \cite{girdhar2023imagebind, zhou2022audio}. \\
        \cmidrule{2-3}
        & High-level Symbolic Abstraction & Converts continuous features into structured symbols via self-supervised learning or 3D geometric-semantic fusion \cite{oquab2023dinov2, miao2023occdepth}. \\
        \midrule
        \multirow{2}{=}{Task-Oriented Environmental Sensing}
        & Goal-Conditioned Selective Attention & Dynamically filters perception streams based on current intent using learned queries or foveal sampling mechanisms \cite{alayrac2022flamingo, zhang2024focus}. \\
        \cmidrule{2-3}
        & Intent-Driven Anomaly Detection & Identifies semantic paradoxes or physical state warnings by checking consistency between perception and logical priors \cite{huang2022inner, le2025rflymad}. \\
        \midrule
        \multirow{2}{=}{Semantic Object Grounding and Tracking}
        & Zero-shot Object Localization & Achieves real-time mapping between linguistic symbols and visual entities in open-world scenarios \cite{liu2024grounding, ravi2024sam}. \\
        \cmidrule{2-3}
        & Spatio-temporal Semantic Consistency & Maintains semantic identity across video frames using reasoning-embedded segmentation or temporal association \cite{lai2024lisa, yang2023track}. \\
        \midrule
        \multirow{2}{=}{Embodied Semantic Environment Understanding} 
        & Semantic Scene Graph Generation & Constructs global topological maps by injecting open-vocabulary semantics into 3D reconstructions \cite{jatavallabhula2023conceptfusion, rosinol20203d}. \\
        \cmidrule{2-3}
        & World Model-based Predictive Perception & Infers future intent and environment evolution using world models or large-scale diffusion architectures \cite{russell2025gaia, wang2024imaginative}. \\
        \bottomrule
    \end{tabular}
\end{table}

Semantics enable the perception stage to accurately identify entities with logical significance and their physical attributes from massive, noisy, and unstructured multi-modal signals. Table \ref{Perception_Stage} illustrates how agents construct a high-dimensional semantic view of the physical world from four dimensions: semantic feature extraction and representation, task-oriented environmental sensing, semantic object grounding and tracking, and embodied semantic environment understanding.


\subsubsection{Semantic Feature Extraction and Representation}
Semantic feature extraction and representation constitute the physical foundation of perception, aiming to map heterogeneous sensor signals into a unified, computable semantic space via deep neural networks, thereby achieving high compression and symbolization of information.

\paragraph{Unified Multi-modal Embedding}
Achieving a unified representation of heterogeneous perceptual information is essential for general cognitive capabilities. To bridge the gaps between modalities, the ImageBind framework proposed by \cite{girdhar2023imagebind} constructs a joint cross-modal embedding space through contrastive learning. By leveraging images as a bridge, it enables agents to achieve emergent semantic alignment and multi-source correlation even without specific paired data. Furthermore, for complex dynamic interactions, Zhou \textit{et al.} \cite{zhou2022audio} design Audio-Visual Segmenter (AVS) to enhance low-level feature fusion of audio-visual modalities. It employs pixel-level synchronous attention to significantly improve spatial localization and semantic binding of sound-producing entities in noisy environments. This unified mechanism ensures that multi-dimensional signals are transformed into compressed symbolic representations without loss of critical information, providing robust input for subsequent decision-making.

\begin{figure}
    \centering
    \includegraphics[width=0.8\linewidth]{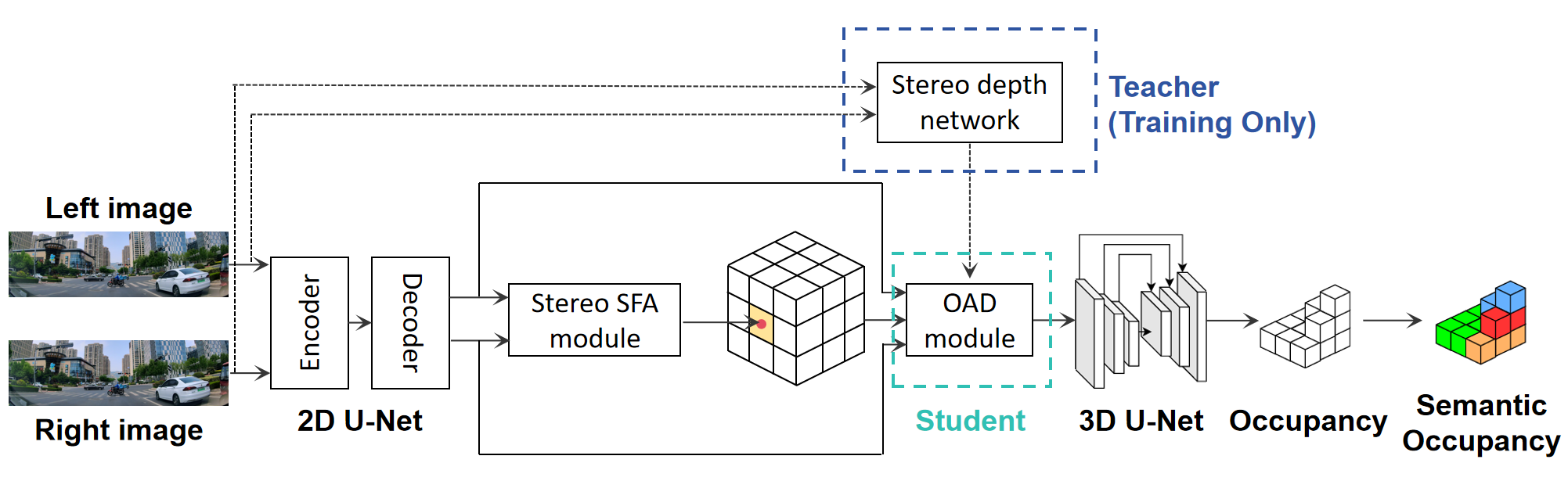}
    \caption{The pipeline of OccDepth for 3D semantic scene completion \cite{miao2023occdepth}. The framework leverages an OAD module for depth enhancement and transforms 2D visual features into 3D semantic representations for environmental perception.}
    \label{fig:OccDepth}
\end{figure}

\paragraph{High-level Symbolic Abstraction}
Following initial feature extraction, agents must transform continuous feature vectors into high-level symbols with logical meaning. In unsupervised learning, the DINOv2 model proposed in \cite{oquab2023dinov2} extracts highly discriminative visual features through large-scale self-supervised pre-training. This scheme overcomes the constraints of human annotation, achieving robust symbolic representation of open-world entities. Additionally, given the importance of spatial dimensions in embodied interaction, Miao \textit{et al.} \cite{miao2023occdepth} develop the OccDepth strategy as illustrated in Figure \ref{fig:OccDepth}. It integrates depth geometry with semantic labels, and this joint modeling facilitates a transition from 2D pixel-level analysis to 3D spatial semantic occupancy symbols. It not only enhances the structural integrity of perception but also provides a solid physical foundation for subsequent autonomous navigation and obstacle avoidance planning in 3D environments.

\subsubsection{Task-Oriented Environmental Sensing}
Emphasizing the subjective initiative of perception, agents dynamically allocate computational resources based on their current intent to achieve selective extraction of key semantic information and conflict detection.

\paragraph{Goal-Conditioned Selective Attention}
In physical environments with high levels of interference, agents must dynamically filter high-value information to reduce cognitive load. The Flamingo architecture proposed in \cite{alayrac2022flamingo} addresses this by introducing a Perceiver Resampler that samples visual features through learned queries via cross-attention. This mechanism enables the agent to concentrate on task-relevant spatial and semantic dimensions, achieving substantial dimensionality reduction.  In contrast, the Focus-Agent framework introduced by Zhang \textit{et al.} \cite{zhang2024focus} draws inspiration from biological vision by emulating the foveal sampling principles of the human retina. It equips agents with intent-driven local fine perception, allowing them to accurately capture high-value details during macro-level logical reasoning with minimal computational cost. This approach effectively mimics human attention allocation strategies in task-specific contexts, significantly enhancing the perceptual autonomy and initiative of agents.

\paragraph{Intent-Driven Anomaly Detection}
Ensuring consistency between environmental information and internal logic is vital, particularly for identifying anomalies that deviate from task intents. The Inner Monologue framework designed by Huang \textit{et al.} \cite{huang2022inner} employs LLMs to perform closed-loop logical reasoning on perceived semantics. By generating continuous internal feedback, the framework monitors conflicts between execution states and intents, identifying logical inconsistencies such as ``missing target objects". For embodied agents, Le \textit{et al.} utilize datasets like RflyMAD \cite{le2025rflymad} to train models that enable unmanned aerial vehicles (UAVs) to perform semantic-level anomaly recognition and collision warning based on kinematic laws. This intent-driven detection enhances self-correction capabilities and elevates perception from passive state recording to active logical verification, marking a significant step toward achieving high autonomy in embodied intelligence.

\subsubsection{Semantic Object Grounding and Tracking}
Semantic object grounding and tracking address the real-time mapping between abstract semantic symbols and concrete physical entities, enabling agents to maintain continuous and consistent attention to specific targets throughout complex mobile interactions.

\paragraph{Zero-shot Object Localization}
A core challenge lies in binding abstract linguistic symbols with concrete physical entities in real time. Grounding DINO proposed in \cite{liu2024grounding} achieves feature alignment between natural language and open-domain visual entities through deep cross-attention mechanisms, allowing agents to precisely locate targets using only text prompts. To address temporal evolution in video streams, Ravi \textit{et al.} \cite{ravi2024sam} propose the SAM 2 that extends zero-shot segmentation to interactive scenarios by incorporating a memory attention module. This enables agents to continuously and accurately segment and track specific entities despite severe deformations, viewpoint changes, or brief occlusions, granting exceptional generalization and adaptability in unknown environments.

\paragraph{Spatio-temporal Semantic Consistency}
In dynamic environments, agents must maintain consistent semantic entity recognition to prevent identity loss due to perspective shifts or occlusions. The LISA framework provided in \cite{lai2024lisa} integrates embedded LLM reasoning to help agents interpret complex implicit instructions and preserve semantic continuity over time, effectively providing a tracking pipeline with ``reasoning memory". Concurrently, Yang \textit{et al.} \cite{yang2023track} design the Track-Anything scheme that combines advanced interactive segmentation with long-term association algorithms. From a visual-spatial perspective, it ensures stable tracking of core semantic entities even in complex scenes involving high-speed motion or non-rigid deformations. Maintaining such spatio-temporal consistency marks the agent's evolution from fragmented instantaneous perception to robust, continuous semantic tracking.

\subsubsection{Embodied Semantic Environment Understanding}
Embodied semantic environment understanding involves the agent's understanding of the overall scene logic, which guides long-term embodied interactive behaviors by constructing topological scene graphs or predicting physical world evolution.

\paragraph{Semantic Scene Graph Generation}
Advanced perception necessitates a shift from isolated entity recognition to macro-topological understanding. In the context of spatial-semantic fusion, ConceptFusion proposed in \cite{jatavallabhula2023conceptfusion} demonstrates significant innovation by injecting open-vocabulary semantics from foundation models into 3D reconstructions at the pixel level. This generates global graphs imbued with spatial logic, allowing agents to ``read" functional attributes and object affordances. 
Furthermore, Rosinol \textit{et al.} \cite{rosinol20203d} propose S-Graph, a bottom-up mechanism for hierarchical scene graph generation. By aggregating local geometric primitives and global topological semantics, it enables agents to autonomously construct multi-level environmental KBs in large-scale complex scenes. Such structured representation serves as a cornerstone for planning long-term, complex tasks.

\paragraph{World Model-based Predictive Perception}
To endow agents with essential foresight, generative world model architectures have been increasingly applied to perception. The GAIA-2 autonomous driving model proposed in \cite{russell2025gaia} identifies physical semantic patterns within massive trajectory data, thereby enabling it to predict the future intents and paths of traffic participants. It significantly enhances safety redundancy. Moreover, inspired by architectures such as Sora, Wang \textit{et al.} \cite{wang2024imaginative} develop a world sensing scheme. Powered by complex diffusion models, it goes beyond understanding static features to simulate global physical evolutions for subsequent moments. This represents a fundamental transformation of perception systems from passive recorders into active predictive engines capable of anticipating physical laws, providing agents with the ability to foresee the future and mitigate risks.

\subsection{Semantic-based Memory Stage}

\begin{table}[!ht]
    \centering
    \caption{Representative approaches for the semantic-based memory stage}
    \label{Memory_Stage}
	\renewcommand{\arraystretch}{1.5}
	\footnotesize
    \begin{tabular}{>{\arraybackslash}m{0.14\textwidth} >{\arraybackslash}m{0.18\textwidth} >{\arraybackslash}m{0.60\textwidth}}
        \toprule
        \textbf{Category} & \textbf{Sub-Category} & \textbf{Descriptions} \\
        \midrule
        \multirow{2}{=}{Hierarchical Semantic Memory Structures}  
        & Working Semantic Memory & Maintains short-term context through recurrent memory mechanisms or high-frequency observation buffers \cite{bulatov2022recurrent, park2023generative}. \\
        \cmidrule{2-3}
        & Long-term Knowledge Consolidation & Integrates experiences across tasks and prevents catastrophic forgetting via generative replay or systematic knowledge banks \cite{chhikara2025mem0, shin2017continual}. \\
        \midrule
        \multirow{2}{=} {Semantic Retrieval and Reasoning}
        & Vector-based Similarity Search & Enables high-speed ANN search in high-dimensional semantic spaces using HNSW indexing and product quantization \cite{malkov2018efficient, johnson2019billion}. \\
        \cmidrule{2-3}
        & Logic-driven Associative Retrieval & Wakes up implicit knowledge through multi-hop KG-RAG reasoning or relational memory-based attention \cite{sanmartin2024kg, santoro2018relational}. \\
        \midrule
        \multirow{2}{=}{Memory Evolution and Knowledge Update}
        & Online Knowledge Base Updating & Synchronizes internal logic with physical reality via incremental scene graphs and LLM-based self-correction \cite{hughes2022hydra, huang2022inner}. \\
        \cmidrule{2-3}
        & Semantic Forgetting and Pruning & Optimizes storage efficiency through utility-based importance scoring or semantic compaction into abstract vectors \cite{park2023generative, wang2023augmenting}. \\
        \midrule
        \multirow{2}{=}{Cognitive Augmentation via Memory} 
        & Contextual Augmentation for Reasoning & Enhances current decisions via self-reflective retrieval-generation or dual instruction tuning \cite{asai2023self, lin2023ra}. \\
        \cmidrule{2-3}
        & Experience-driven Planning & Optimizes future action sequences through failure-based self-reflection or reasoning-as-planning search \cite{shinn2023reflexion, hao2023reasoning}. \\
        \bottomrule
    \end{tabular}
\end{table}

The memory stage enables agents to break information silos and achieve lifelong continuous learning. By transforming transient high-dimensional information briefly captured in the perception stage into a highly structured, long-term queryable underlying KB, semantic-based memory provides solid prior background support for agents' long-term planning decisions. 
As illustrated in Table \ref{Memory_Stage}, we summarize representative approaches for the semantic-based memory stage, including hierarchical semantic memory structures, semantic retrieval and reasoning, memory evolution and knowledge update, and cognitive augmentation via memory.


\subsubsection{Hierarchical Semantic Memory Structures}
The semantic-based memory is systematically divided into working semantic memory for immediate logical processing and long-term knowledge consolidation for persistent learning.

\paragraph{Working Semantic Memory}
Working semantic memory serves as the agent's ``cache" during active tasks, maintaining short-term contextual consistency. To mitigate information loss in long sequences, the RMT architecture proposed in \cite{bulatov2022recurrent} introduces recursive memory tokens, enabling agents to retain critical transient semantics across sequences of up to a million tokens. Additionally, for high-frequency semantic fragments, Park \textit{et al.} \cite{park2023generative} develop the memory stream buffer, which offers an efficient experiential buffering mechanism. This mimics biological short-term memory, allowing agents to extract features directly from the cache during complex interactions without repeatedly accessing long-term storage, thereby enabling low-latency responses. This hierarchical caching significantly improves both reaction speed and logical consistency in real-time tasks.

\begin{figure}
    \centering
    \includegraphics[width=0.8\linewidth]{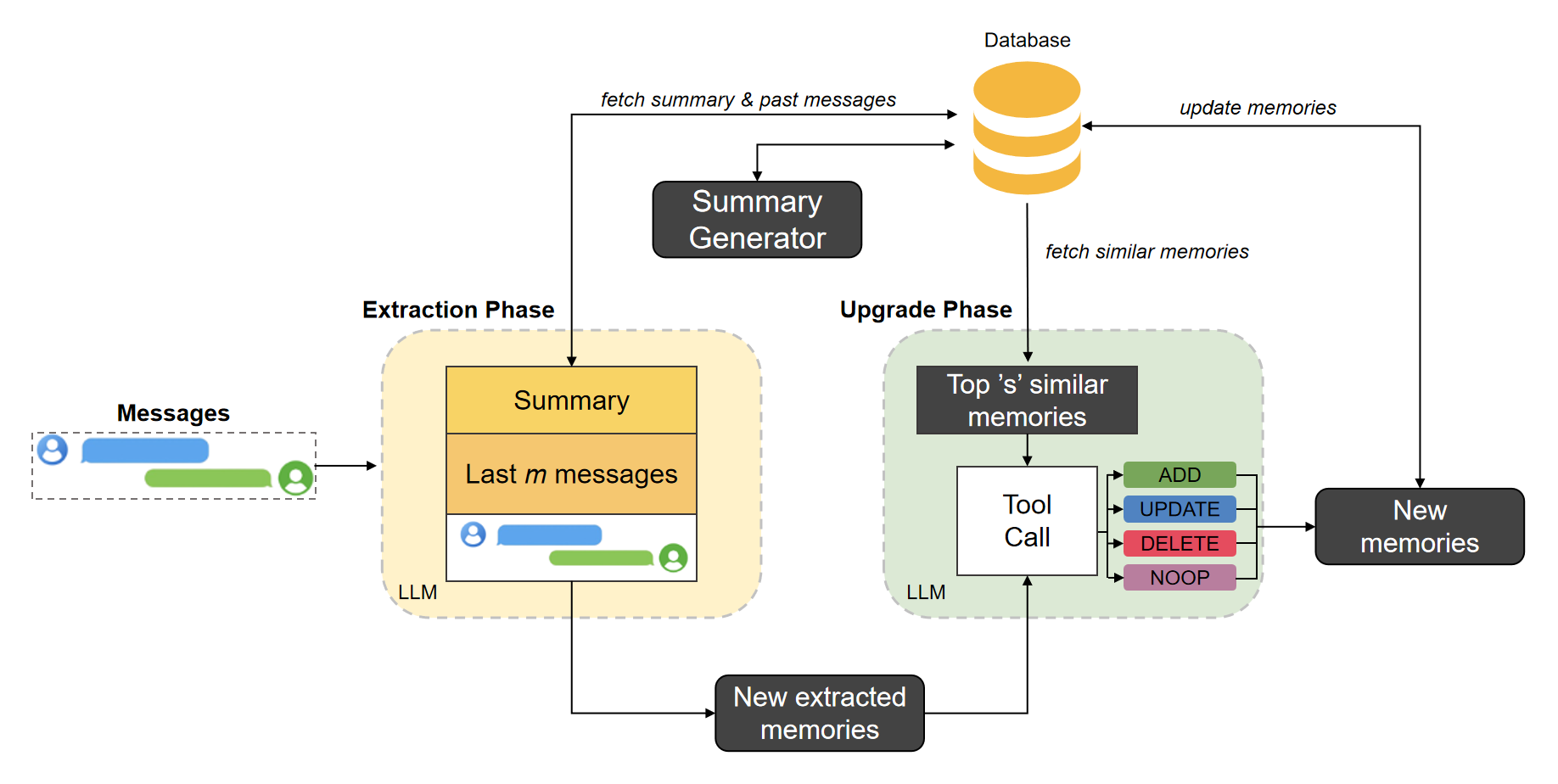}
    \caption{Architectural overview of the Mem0 system \cite{chhikara2025mem0}, illustrating the semantic extraction and memory update phases. The system processes historical context to generate new memories, which are then evaluated against existing records and refined via a Tool Call mechanism before being stored in the central database.}
    \label{fig:Mem0}
\end{figure}

\paragraph{Long-term Knowledge Consolidation}
Long-term memory transforms fragmented experiences accumulated over the agent's lifecycle into systematic common sense. As illustrated in Figure \ref{fig:Mem0}, the Mem0 framework proposed in \cite{chhikara2025mem0} represents a notable advance in this direction, enabling robust cross-session and cross-task memory integration that builds personalized semantic assets over time. To address the challenge of catastrophic forgetting, Shin \textit{et al.} \cite{shin2017continual} design the deep generative replay architecture, which trains a ``scholar" model to replay historically valuable semantic information. This allows agents to consolidate existing knowledge while acquiring new skills without relying on stored raw data. This approach balances knowledge acquisition and retention, enabling the agent to evolve from a simple executor into a lifelong learner.

\subsubsection{Semantic Retrieval and Reasoning}
Semantic retrieval and reasoning investigate how agents accurately extract highly relevant knowledge from vast historical memory repositories, thereby achieving on-demand instant activation.


\paragraph{Vector-based Similarity Search}
For engineering implementation of massive unstructured knowledge matching, agents rely heavily on efficient vector retrieval. Malkov \textit{et al.} \cite{malkov2018efficient} propose the Hierarchical Navigable Small World (HNSW) algorithm, which employs multi-layer graph indices to enable fast Approximate Nearest Neighbors Search (ANNS) in billion-scale vector spaces, effectively breaking latency bottlenecks in long-term memory access. 
In large-scale deployment, the Faiss framework proposed in \cite{johnson2019billion} further reduces memory footprint through GPU-parallel computing and product quantization, while also incorporating intent-weighting strategies to enhance retrieval intelligence. This ensures that retrieved historical experiences are not only mathematically similar to queries but also logically aligned with the current functional execution context. Such efficient retrieval serves as the technological foundation for on-demand knowledge activation.

\paragraph{Logic-driven Associative Retrieval}
Unlike simple spatial distance comparisons, associative retrieval emphasizes deep logical connections between knowledge fragments. Sanmartin \textit{et al.} \cite{sanmartin2024kg} integrates KG topologies with retrieval-augmented generation (RAG), enabling agents to perform multi-hop reasoning along semantic edges and retrieve latent knowledge points. Similarly, relational memory networks proposed in \cite{santoro2018relational} construct dynamic associative matrices via self-attention mechanisms. This approach avoids fixed physical memory addressing, instead computing interactions between memory units in real time based on attention. It simulates the brain's non-linear associative process—driven by semantic relatedness rather than physical distance. This logic-driven mechanism enhances retrieval precision and equips agents with superior common-sense integration and associative reasoning capabilities in complex contexts.

\subsubsection{Memory Evolution and Knowledge Update}
Memory systems must function as dynamic, living structures rather than static repositories. Memory evolution and knowledge update examine how agents maintain accuracy, timeliness, and retrieval efficiency in large-scale KBs under limited computational resources, achieved through continuous online streaming learning and selective pruning of redundant information.

\paragraph{Online Knowledge Base Updating}
Given the rapid changes in physical environments, agents must continuously adjust internal logical structures to maintain relevance. Hughes \textit{et al.} \cite{hughes2022hydra} propose the Hydra framework, which enables robots to dynamically update semantic nodes and weights based on streaming data without triggering global restructuring, ensuring the KB remains synchronized with real-world conditions. 
When perceptual input conflicts with existing experience, the Inner Monologue mechanism proposed in \cite{huang2022inner} provides a closed-loop error correction tool. 
It leverages LLMs as logical discriminators to evaluate perceptual confidence against memory consistency, autonomously correcting erroneous entries and iterating obsolete task knowledge. This online evolution enables agents to flexibly adapt to environmental changes and achieve real-time cognitive updates during long-term deployment.

\paragraph{Semantic Forgetting and Pruning}
To prevent memory overload caused by redundant information, intelligent semantic forgetting and compression mechanisms are essential.
The generative agent framework proposed in \cite{park2023generative} employs a utility-decay pruning model that scores semantic fragments based on importance, relevance, and recency, thereby removing low-value data and maintaining a lightweight KB.
Alternatively, the LongMem architecture proposed in \cite{wang2023augmenting} explores semantic compaction by using memory-augmented Transformers to encode concrete experiences into abstract, compact knowledge vectors. This preserves core logical content while significantly reducing computational overhead and cache burden during long-term storage and retrieval. Together, these pruning and compression mechanisms draw inspiration from biological forgetting principles, substantially enhancing agent efficiency in processing large-scale, long-duration tasks.

\subsubsection{Cognitive Augmentation via Memory}
By seamlessly integrating historical experiences into current reasoning pipelines and future execution blueprints, agents can achieve highly context-aware and foresighted decision-making capabilities.

\paragraph{Contextual Augmentation for Reasoning}
The core function of contextual augmentation is to enhance ongoing reasoning and decision-making.
The Self-RAG framework proposed by Asai \textit{et al.} \cite{asai2023self} introduces ``reflection tokens," enabling agents to autonomously determine when retrieval is needed, assess the quality of retrieved information, and ensure factual accuracy. Thus, it improves decision rigor and self-consistency.
Building on this, Lin \textit{et al.} \cite{lin2023ra} proposed RA-DIT to enable agents to integrate background knowledge to resolve perceptual ambiguity. This allows agents to disambiguate incomplete or noisy perceptual inputs through semantic reasoning, arriving at globally optimal judgments. This shift from passive perception to augmented reasoning renders agents more robust and rational when confronting unknown challenges.

\paragraph{Experience-driven Planning}
By learning from historical successes and failures, agents can achieve foresighted long-term task planning. The Reflexion architecture proposed in \cite{shinn2023reflexion}  endows agents with self-reflection capabilities, enabling them to analyze past failures and optimize future actions without human intervention, thereby avoiding repeated errors.
Additionally, heuristic search planners such as RAP proposed in \cite{hao2023reasoning}, transform optimal memory sequences into internal world models. By leveraging experienced LLMs as sandboxes, they simulate state transitions within memory to distill optimal paths, which then serve as templates for generating new complex tasks. This experience-driven planning establishes a closed-loop learning process, significantly enhancing both execution efficiency and success rates in complex physical tasks.

\subsection{Semantic-based Reasoning Stage}

In the reasoning stage, semantics enable agents to perform analysis, planning, and logical deduction by integrating perceived information with stored knowledge to produce actionable decisions. 
As illustrated in Table~\ref{tab:reasoning}, we review representative approaches across five dimensions: CoT reasoning, KG-augmented reasoning, retrieval-augmented reasoning, tree-structured multi-path reasoning, and neuro-symbolic reasoning.

\begin{table}[!ht]
    \centering
    \caption{Representative approaches for the semantic-based reasoning stage}
    \label{tab:reasoning}
    \renewcommand{\arraystretch}{1.5}
    \footnotesize
    \begin{tabular}{>{\arraybackslash}m{0.14\textwidth} >{\arraybackslash}m{0.23\textwidth} >{\arraybackslash}m{0.55\textwidth}}
        \toprule
        \textbf{Category} & \textbf{Sub-Category} & \textbf{Descriptions} \\
        \midrule
        \multirow{2}{=}{Chain-of-Thought Reasoning}
        & Step Decomposition & Elicits step-by-step intermediate reasoning from LLMs~\cite{wei2022,kojima2022zeroshot}; least-to-most prompting further enables progressive sub-problem decomposition without task-specific exemplars~\cite{zhou2023leasttomost}. \\
        \cmidrule{2-3}
        & Path Voting & Samples diverse reasoning paths and selects answers by majority vote~\cite{wang2023consistency}; complexity-based prompting biases sampling toward higher-quality reasoning chains~\cite{fu2023complexity}. \\
        \midrule
        \multirow{2}{=}{KG-Augmented Reasoning}
        & Graph-LM Fusion & Fuses LM representations with GNN propagation over KG subgraphs~\cite{yasunaga2021}; GreaseLM extends this with multi-layer interleaved LM-graph fusion~\cite{zhang2022greaselm}. \\
        \cmidrule{2-3}
        & Iterative KG Query & Decomposes queries into sequential structured reading operations over KGs~\cite{jiang2023structgpt}; Graph of Thoughts further enables non-linear multi-path reasoning~\cite{besta2024got}. \\
        \midrule
        \multirow{2}{=}{Retrieval-Augmented Reasoning}
        & Retrieval-Conditioned Generation & Combines parametric generation with dense dual-encoder retrieval over large document indices to inject up-to-date external knowledge~\cite{lewis2020rag,karpukhin2020dpr}. \\
        \cmidrule{2-3}
        & Multi-Doc Fusion & Pre-trains retrieval end-to-end with the language model~\cite{guu2020realm}; Fusion-in-Decoder aggregates evidence from multiple retrieved passages at the decoder~\cite{izacard2021fid}. \\
        \midrule
        \multirow{2}{=}{Tree-Structured Multi-Path Reasoning}
        & Tree Search & Frames reasoning as deliberate tree search with backtracking~\cite{yao2023tot}; LATS integrates MCTS with real-environment observations for dynamic replanning~\cite{zhou2024lats}. \\
        \cmidrule{2-3}
        & Process Reward & Assigns dense step-level scores to intermediate reasoning steps~\cite{lightman2024}; process-based feedback outperforms outcome supervision as solution complexity grows~\cite{uesato2022process}. \\
        \midrule
        \multirow{2}{=}{Neuro-Symbolic Reasoning}
        & Program-Aided Reasoning & Offloads computation to a program interpreter~\cite{gao2023pal}; Program of Thoughts further decouples multi-step numerical reasoning from semantic parsing~\cite{chen2023pot}. \\
        \cmidrule{2-3}
        & Logic-Symbolic Execution & Translates problems into formal logical representations executed by symbolic solvers~\cite{pan2023logiclm}; NS-CL jointly learns visual concepts and symbolic reasoning under weak supervision~\cite{mao2019nscl}. \\
        \bottomrule
    \end{tabular}
\end{table}

\subsubsection{Chain-of-Thought Reasoning}

CoT reasoning enables agents to generate step-by-step intermediate reasoning, thereby decomposing complex problems into manageable components. This approach significantly enhances performance on tasks that require multi-step logical deduction. Representative methods include step decomposition and path voting.

\paragraph{Step Decomposition}
Standard CoT prompting improves LLM reasoning by instructing models to articulate intermediate reasoning steps prior to delivering a final answer. Wei \textit{et al.} \cite{wei2022} demonstrate that simply prompting models to reason step by step yields substantial performance gains across arithmetic, commonsense, and symbolic reasoning benchmarks, with the most pronounced improvements observed at larger model scales. Extending this paradigm to zero-shot settings, Kojima \textit{et al.}~\cite{kojima2022zeroshot} show that appending the phrase ``Let’s think step by step” suffices to elicit robust multi-step reasoning without task-specific exemplars, establishing structured intermediate reasoning as a general and transferable capability. Furthermore, Zhou \textit{et al.}~\cite{zhou2023leasttomost} introduce least-to-most prompting, which decomposes a problem into a sequence of progressively simpler sub-problems and solves them in order. 

\paragraph{Path Voting}
Single-path CoT is susceptible to the brittleness of greedy decoding, wherein an early error propagates and amplifies through subsequent steps. Wang \textit{et al.}~\cite{wang2023consistency} address this limitation by sampling diverse reasoning paths and selecting the final answer via majority vote. This approach achieves performance improvements exceeding 17 percentage points over greedy CoT on benchmarks such as GSM8K and StrategyQA. To further enhance the quality of sampled reasoning chains prior to voting, Fu \textit{et al.}~\cite{fu2023complexity} propose complexity-based prompting, which biases sampling toward chains with greater reasoning depth and substantially outperforms uniform random sampling on multi-step benchmarks. 


\subsubsection{KG-Augmented Reasoning}

KG-augmented reasoning grounds inference in structured relational knowledge, allowing agents to reason over factual information that extends beyond the static knowledge encoded in model parameters. When agents must operate under rapidly evolving conditions, such as real-time network topology changes or emerging cross-domain task requirements, access to external structured knowledge becomes essential for maintaining reasoning accuracy and consistency.

\begin{figure}
    \centering
    \includegraphics[width=0.8\linewidth]{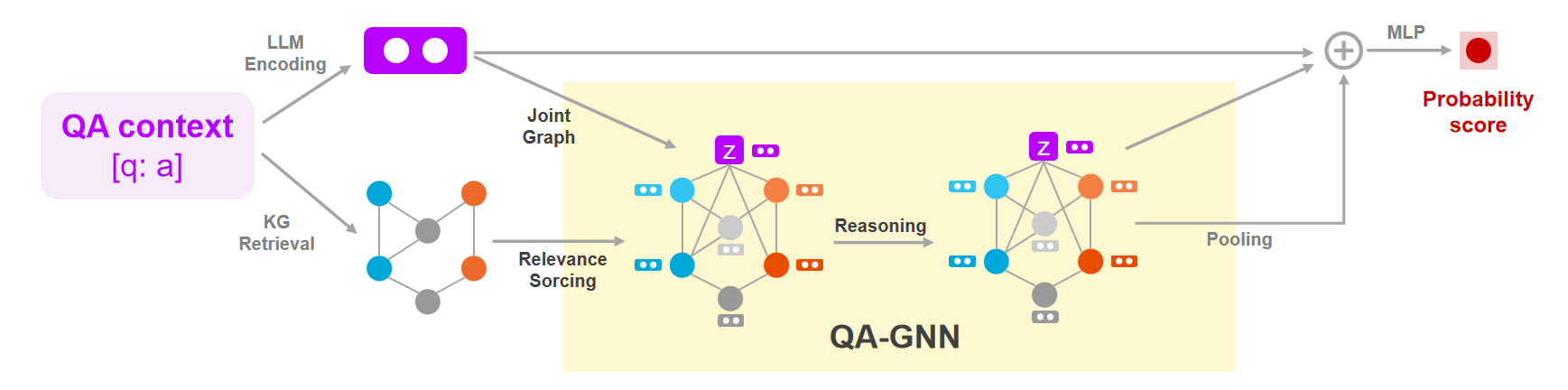}
    \caption{Overview of the QA-GNN reasoning framework\cite{yasunaga2021}, illustrating the joint graph construction by connecting the LLM-encoded QA context with the retrieved knowledge graph to perform context-conditioned reasoning and probability scoring.}
    \label{fig:QA-GNN}
\end{figure}

\paragraph{Graph-LM Fusion}
Graph-LM integrates language model (LM) representations with graph neural network (GNN) propagation to better ground textual reasoning in structured KGs. As illustrated in Figure \ref{fig:QA-GNN}, Yasunaga \textit{et al.}~\cite{yasunaga2021} construct a working graph by linking question-answer entities to relevant KG subgraphs and perform bidirectional message passing between LM contextual representations and GNN node embeddings. This allows relational structure to modulate text-based reasoning directly, rather than treating the KG as a static lookup table. Extending this integration to achieve tighter cross-modal fusion, Zhang \textit{et al.}~\cite{zhang2022greaselm} propose GreaseLM, which inserts dedicated interaction layers at multiple intermediate encoder phases. This design enables structured relational context to progressively refine token representations throughout encoding, yielding consistent improvements on commonsense and biomedical reasoning benchmarks.

\paragraph{Iterative KG Query}
Complex queries often require synthesizing information across multiple heterogeneous structured sources, making iterative interaction essential. Jiang \textit{et al.}~\cite{jiang2023structgpt} introduce StructGPT, which provides LLMs with a unified interface of specialized reading functions covering KGs, relational tables, and document databases. The framework decomposes complex queries into sequential sub-operations that alternately read from these sources and update the reasoning state. 
To further support non-linear reasoning, Besta \textit{et al.}~\cite{besta2024got} propose the Graph of Thoughts (GoT) framework, which represents reasoning outputs as nodes and their logical interdependencies as directed edges in an arbitrary graph topology, enabling aggregation of partial results, iterative node refinement, and non-linear backtracking for complex multi-agent task planning.

\subsubsection{Retrieval-Augmented Reasoning}

Retrieval-augmented reasoning addresses the scalability limitations of static KGs by retrieving relevant evidence on demand from large unstructured corpora, enabling agents to remain current with continuously expanding information environments. 


\paragraph{Retrieval-Conditioned Generation}
RAG conditions a parametric sequence-to-sequence model on retrieved passages to inject external knowledge into the generation process. Lewis \textit{et al.}~\cite{lewis2020rag} formalize this paradigm by retrieving top-$k$ passages from a dense document index and conditioning generation on these results, achieving state-of-the-art performance on multiple open-domain question answering benchmarks while retaining the ability to update knowledge without full model retraining. To substantially improve retrieval quality, Karpukhin \textit{et al.}~\cite{karpukhin2020dpr} introduce Dense Passage Retrieval (DPR), which replaces sparse Term Frequency-Inverse Document Frequency (TF-IDF) retrieval with dense dual-encoder representations trained on question-passage pairs via contrastive learning, consistently outperforming BM25 baselines by large margins. 


\paragraph{Multi-Doc Fusion}
A fundamental limitation of post-hoc retrieval is that model representations are not inherently trained to benefit from retrieved evidence. Guu \textit{et al.}~\cite{guu2020realm} address this gap through Retrieval-Augmented Language Model pre-training (REALM), which embeds a latent knowledge retriever directly into the masked language model pre-training objective. This approach trains the retriever end-to-end alongside the language model backbone, jointly optimizing representations for both language understanding and knowledge retrieval. 
For settings requiring multi-document evidence synthesis, Izacard and Grave~\cite{izacard2021fid} propose FiD, which processes each retrieved passage independently through a shared encoder and fuses all representations collectively at the decoder via cross-attention. It substantially outperforms single-passage conditioning and establishes a robust evidence integration mechanism for complex semantic reasoning tasks.

\subsubsection{Tree-Structured Multi-Path Reasoning}

Tree-structured multi-path reasoning reformulates reasoning as a deliberate tree search, overcoming the fundamental limitation of linear reasoning: each step commits irrevocably to a single direction with no mechanism for recovery from early errors. 


\paragraph{Tree Search}
Tree of Thoughts (ToT) enables agents to explore multiple competing reasoning paths through systematic tree search. Yao \textit{et al.}~\cite{yao2023tot} design a framework in which, at each node, the model generates multiple candidate thought continuations, evaluates their promise using self-evaluation heuristics, and applies Breadth-First Search (BFS) or Depth-First Search (DFS) to navigate the reasoning space, substantially outperforming both standard CoT and self-consistency decoding, with the performance gap widening as problem difficulty increases. To unify tree-structured reasoning with interactive environment feedback, Zhou \textit{et al.}~\cite{zhou2024lats} propose Language Agent Tree Search (LATS), which integrates Monte Carlo Tree Search (MCTS) with LLM-based state evaluation and real-world observations. LATS enables systematic backtracking, lookahead value estimation, and dynamic replanning in response to execution outcomes. 


\paragraph{Process Reward}
Tree search introduces a critical challenge: unreliable heuristics for evaluating candidate branches can lead to systematic exploration of low-quality subtrees. Lightman \textit{et al.}~\cite{lightman2024} address this through Process Reward Models (PRMs), which assign dense scalar scores to individual intermediate reasoning steps. Through large-scale human annotation, they demonstrate that step-level supervision substantially outperforms outcome-supervised models in best-of-N candidate selection. Offering a complementary perspective, Uesato \textit{et al.}~\cite{uesato2022process} systematically compare process-based and outcome-based feedback, showing that process-level supervision delivers a more precise and sample-efficient training signal, with the performance gap widening as solution complexity increases. 


\subsubsection{Neuro-Symbolic Reasoning}

Neuro-symbolic reasoning addresses the inherent unreliability of token-level computation for arithmetic and logical tasks by delegating formal computation to deterministic external systems while retaining LLMs for natural language comprehension and high-level semantic understanding. 


\paragraph{Program-Aided Reasoning}
Program-Aided Reasoning (PAL) separates semantic understanding from formal computation by offloading symbolic operations to a deterministic program interpreter. Gao \textit{et al.}~\cite{gao2023pal} instruct LLMs to translate natural-language reasoning problems into executable Python programs, achieving state-of-the-art results on mathematical and symbolic reasoning benchmarks by eliminating arithmetic errors and logic inversions that cannot be resolved through scaling or prompting alone. Extending this to more expressive computation structures, Chen \textit{et al.}~\cite{chen2023pot} propose Program of Thoughts (PoT) prompting, which generates programs containing explicit iterative and recursive computation patterns to decouple complex multi-step arithmetic from semantic parsing, achieving further gains on financial, scientific, and mathematical benchmarks. 


\paragraph{Logic-Symbolic Execution}
While program synthesis effectively handles numerical computation, it does not enforce logical consistency across multi-step relational reasoning. Pan \textit{et al.}~\cite{pan2023logiclm} propose Logic-LM, which translates natural-language problems into formal logical representations encompassing first-order logic and constraint satisfaction, submitting them to dedicated symbolic solvers for guaranteed-correct execution. An LLM-based self-refinement loop iteratively corrects translation errors based on solver feedback. Complementing this scheme, Mao \textit{et al.}~\cite{mao2019nscl} propose Neuro-Symbolic Concept Learner (NS-CL), which jointly learns visual concepts, semantic language parsing, and symbolic program execution within a unified architecture under weak supervision, achieving strong systematic generalization by disentangling perceptual primitive learning from compositional reasoning.

\subsection{Semantic-based Action Stage}

Traditional action methods are heavily reliant on rigid API formats, where even minor deviations in instruction can cause system failure. In contrast, semantic-enhanced agents understand user intent and automatically map it to the required API parameters. Moreover, if an error occurs during execution, semantic analysis can diagnose the cause and adjust the strategy in real time. As shown in Table~\ref{tab:action}, we evaluate representative approaches across five dimensions: semantic tool acquisition, reasoning-action interleaving, multi-agent collaborative action, semantic self-correction, and reinforcement-based semantic feedback.

\begin{table}[!ht]
    \centering
    \caption{Representative approaches for the semantic-based action stage}
    \label{tab:action}
    \renewcommand{\arraystretch}{1.5}
    \footnotesize
    \begin{tabular}{>{\arraybackslash}m{0.14\textwidth} >{\arraybackslash}m{0.23\textwidth} >{\arraybackslash}m{0.55\textwidth}}
        \toprule
        \textbf{Category} & \textbf{Representative Schemes} & \textbf{Descriptions} \\
        \midrule
        \multirow{2}{=}{Semantic Tool Acquisition}
        & Toolformer & Trains LLMs to decide autonomously when and how to call external tools via self-supervised annotation filtering, without large-scale human annotation~\cite{schick2023toolformer}. \\
        \cmidrule{2-3}
        & ToolLLM & Constructs a repository of 16,000+ real-world REST APIs and trains LLMs via DFSDT inference to invoke previously unseen APIs through instruction tuning~\cite{qin2024toolllm}. \\
        \midrule
        \multirow{2}{=}{Reasoning-Action Interleaving}
        & ReAct & Interleaves verbal reasoning traces and external environment actions within the LLM output stream, enabling dynamic re-planning based on real-time observation feedback~\cite{yao2023react}. \\
        \cmidrule{2-3}
        & HuggingGPT & Uses an LLM as a controller to parse requests, select expert models from a model hub, execute them in dependency order, and synthesize results for complex multi-modal tasks~\cite{shen2023hugginggpt}. \\
        \midrule
        \multirow{2}{=}{Multi-Agent Collaborative Action}
        & MetaGPT & Encodes software engineering workflows as structured SOPs and assigns role-specific agents to execute them collaboratively, substantially reducing incoherence in multi-LLM outputs~\cite{hong2024metagpt}. \\
        \cmidrule{2-3}
        & Generative Agents & Equips LLM-driven agents with persistent memory streams, reflection synthesis, and retrieval-based planning, enabling emergent social behaviors across extended multi-agent interactions~\cite{park2023generative}. \\
        \midrule
        \multirow{2}{=}{Semantic Self-Correction}
        & Reflexion & Reinforces language agents via verbal self-assessment stored in episodic memory, prepended to subsequent task contexts without any gradient updates~\cite{shinn2023reflexion}. \\
        \cmidrule{2-3}
        & Self-Refine & Employs a single LLM as generator, critic, and refiner in an iterative loop, consistently improving outputs across diverse tasks without additional training~\cite{madaan2023selfrefine}. \\
        \midrule
        \multirow{2}{=}{Reinforcement-Based Semantic Feedback}
        & InstructGPT & Aligns LLMs with human preferences via supervised fine-tuning on demonstrations, reward model training on pairwise rankings, and PPO optimization~\cite{ouyang2022instructgpt}. \\
        \cmidrule{2-3}
        & Recursive Reward Modeling & Applies RLHF to long-form summarization, demonstrating that learned reward models can supervise tasks exceeding direct human evaluation capacity~\cite{stiennon2020}. \\
        \bottomrule
    \end{tabular}
\end{table}

\subsubsection{Semantic Tool Acquisition}

Semantic tool acquisition enables agents to autonomously determine when and how to invoke external tools, replacing manually crafted rules with learned tool selection and invocation.

\paragraph{Toolformer}
Toolformer \cite{schick2023toolformer} allows LLMs to learn when and how to call external tools through self-supervised learning. It first prompts the model to leverage candidate API call insertions within generated text, then retain only those calls whose returned results measurably reduce language modeling loss on the surrounding context. Fine-tuning on this self-filtered dataset teaches the model to recognize when retrieval is genuinely beneficial, rather than merely syntactically plausible. Evaluated across calculators, calendars, search engines, and translation APIs, Toolformer achieves competitive zero-shot performance without task-specific training. 

\begin{figure}
    \centering
    \includegraphics[width=0.8\linewidth]{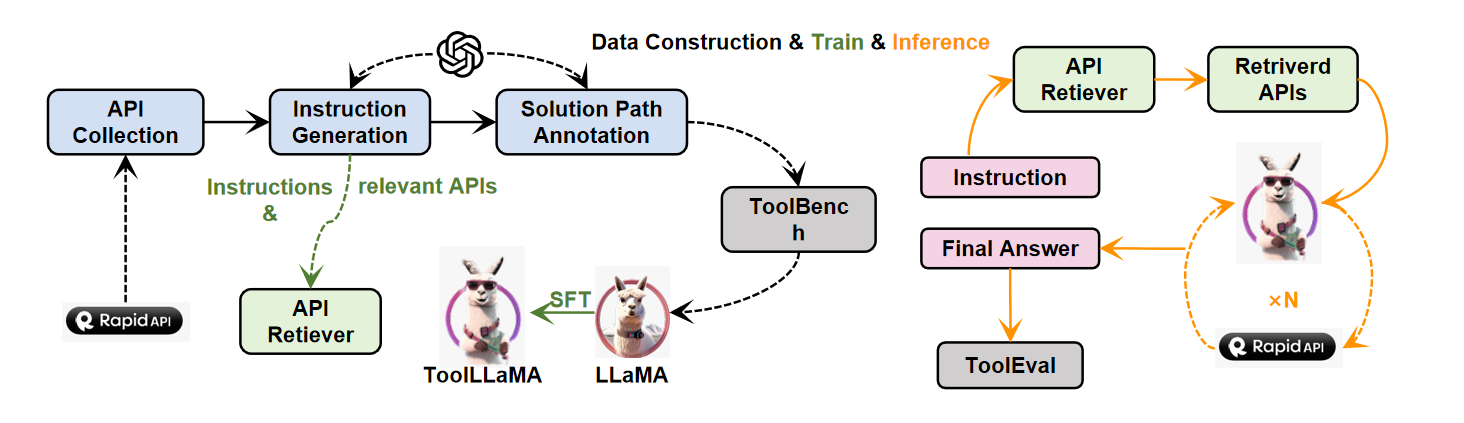}
    \caption{Overview of the ToolBench pipeline\cite{qin2024toolllm}, illustrating the three-phase construction process of data annotation and the dual-stage inference workflow where an API retriever provides context for ToolLLaMA to execute multi-round reasoning.}
    \label{fig:ToolLLaMA}
\end{figure}

\paragraph{ToolLLM}
Toolformer is trained on a small, curated set of APIs and does not scale to the breadth of tools encountered in real-world deployments. As depicted in Figure \ref{fig:ToolLLaMA}, Qin \textit{et al.} \cite{qin2024toolllm} address this scalability challenge by constructing ToolBench, a large-scale instruction-tuning dataset spanning over 16,000 real-world REST APIs across 49 categories. To handle the combinatorial difficulty of selecting and sequencing calls across this tool space, ToolLLM introduces a depth-first search-based decision tree (DFSDT) inference strategy that explores multiple tool-call paths and backtracks from failed branches, an explicit search mechanism that mirrors the tree-structured reasoning. The resulting model generalizes effectively to unseen APIs at inference time. 
In semantic-based agent communication networks serving heterogeneous vertical industries, this zero-shot generalization capability is essential for application agents interfacing with industry-specific network management APIs without requiring per-API fine-tuning.

\subsubsection{Reasoning-Action Interleaving}

Reasoning-action interleaving dissolves the boundary between internal planning and external execution, allowing agents to dynamically replan based on real-time environmental feedback.

\paragraph{ReAct}
ReAct \cite{yao2023react} interleaves verbal reasoning traces and environment-facing actions within the same token stream, enabling dynamic replanning in response to new observations. The model alternates between producing natural-language reasoning steps and issuing tool calls or environment queries, with each observation immediately informing the next reasoning step. Simulation results show that, ReAct substantially outperforms both reasoning-only and acting-only baselines, and the transparency of its interleaved traces facilitates straightforward failure diagnosis. 


\paragraph{HuggingGPT}
While ReAct operates as a single-agent system, its throughput is limited when tasks require the coordination of multiple specialized capabilities. Shen \textit{et al.}~\cite{shen2023hugginggpt} address this limitation by introducing HuggingGPT, which employs an LLM as a meta-controller. The controller decomposes user requests into structured task plans, selects the most appropriate expert model from a model hub for each subtask, executes them in dependency order, and synthesizes their outputs into a unified response. This framework effectively handles complex cross-modal tasks involving vision, speech, video, and language that no single model can manage alone.

\subsubsection{Multi-Agent Collaborative Action}

Multi-agent collaborative action shifts the fundamental unit of action from a single agent to a coordinated team, addressing the coherence challenge of ensuring that multiple agents working in parallel produce mutually consistent outputs and collectively advance toward shared goals.

\paragraph{MetaGPT}
MetaGPT \cite{hong2024metagpt} encodes established human collaborative workflows as structured communication protocols that govern role-specific agent behavior. Specifically, it formalizes software engineering standard operating procedures (SOPs) into structured message schemas: a product manager agent decomposes requirements, an architect agent produces system designs, engineer agents implement components, and a tester agent validates outputs, with all communicating through structured schemas rather than free-form natural language. Simulation results show that, MetaGPT substantially reduces hallucination and incoherence compared to unconstrained multi-agent baselines.

\paragraph{Generative Agents}
While MetaGPT relies on pre-specified protocols tailored to a specific task domain, enabling coherent collaborative behavior over extended timescales without such protocols presents a more general challenge. Park \textit{et al.}~\cite{park2023generative} address this through a sandbox simulation in which 25 LLM-driven agents interact over an extended period. 
Each agent is equipped with a memory stream that logs experiences as natural-language observations, a reflection mechanism that periodically synthesizes high-level insights from retrieved memories, and a planning module that translates those insights into behavioral intentions. Despite the absence of explicit coordination instructions, the agents exhibit emergent relationship formation, information propagation, and collaborative event organization.

\subsubsection{Semantic Self-Correction}

Semantic self-correction enables agents to diagnose execution failures, extract generalizable lessons, and improve subsequent attempts without modifying model weights, thereby supporting autonomous operation in resource-constrained environments.

\paragraph{Reflexion}
Reflexion \cite{shinn2023reflexion} introduces a form of verbal reinforcement learning that enables agent self-improvement without gradient updates. After each task attempt, the agent generates a written self-assessment identifying failure points and improvement strategies, stores this reflection in an episodic memory buffer, and prepends it to the context in subsequent attempts. On AlfWorld, HotpotQA, and HumanEval, Reflexion substantially outperforms baseline agents. Besides, on programming tasks, it approaches the performance of GPT-4. 

\paragraph{Self-Refine}
While Reflexion addresses failures at the episode level space,many tasks benefit from fine-grained correction within a single execution. Against this background, Madaan \textit{et al.}~\cite{madaan2023selfrefine} propose Self-Refine, a scheme that employs a single LLM in three distinct roles within an iterative loop: generator, critic, and refiner. Given an initial output, the model critiques it along task-relevant dimensions and revises it accordingly, continuing until a termination condition is met. Evaluated across seven tasks including code generation, text rewriting, and mathematical reasoning, Self-Refine consistently improves over single-pass baselines without additional training. 


\subsubsection{Reinforcement-Based Semantic Feedback}

Reinforcement-based semantic feedback offers a training-time mechanism for aligning agent behavior with target objectives, complementing the inference-time self-correction approaches discussed above.

\paragraph{InstructGPT}
InstructGPT \cite{ouyang2022instructgpt} demonstrates the effectiveness of aligning LLMs with human preferences through a three-stage pipeline. It combines supervised fine-tuning on human demonstrations, reward model training on pairwise preference rankings, and proximal policy optimization (PPO), producing models that human evaluators substantially prefer over much larger unaligned models across a wide range of tasks. The reward model acts as a learned proxy for human judgment, capturing evaluation criteria that are difficult to specify as explicit loss functions. 


\paragraph{Recursive Reward Modeling}
While InstructGPT operates on short-form tasks where human evaluation is relatively tractable, the challenging application lies in tasks whose outputs exceed direct human assessment capacity. To address this, Stiennon \textit{et al.}~\cite{stiennon2020} train a reward model on pairwise human preferences over summaries and optimize a policy via PPO against this learned reward. Simulation results verify that a learned reward model can provide reliable training signal for tasks exceeding direct evaluation capacity.


\section{AI Agents for SemCom Networks}
\label{sectionagent}

This section outlines four representative types of AI agents for SemCom networks: embodied agents, communication agents, network agents, and application agents. Table \ref{agenttype} illustrates representative AI agents.

\begin{table}[!ht]
    \centering
    \caption{Representative AI agents for SemCom networks}
    \label{agenttype}
    \renewcommand{\arraystretch}{1.5} 
    \footnotesize
    \setlength{\tabcolsep}{2pt}
    
    \begin{tabular}{>{\centering\arraybackslash}m{0.14\textwidth} >{\centering\arraybackslash}m{0.12\textwidth} >{\raggedright\arraybackslash}m{0.22\textwidth} >{\raggedright\arraybackslash}m{0.22\textwidth} >{\centering\arraybackslash}m{0.12\textwidth} >{\centering\arraybackslash}m{0.10\textwidth}}
    \toprule
    \textbf{Category} & \multicolumn{2}{c}{\textbf{Project}} & \textbf{Company} & \textbf{Release Date} & \textbf{Open Source} \\
    \midrule
    
    \multirow{2}{=}{\centering Embodied Agent} & 
    \multicolumn{2}{>{\centering\arraybackslash}m{0.34\textwidth}}{SayCan} & Google & 2022 & \checkmark \\
    \cmidrule{2-6}
    & \multicolumn{2}{>{\centering\arraybackslash}m{0.34\textwidth}}{Atlas} & Boston Dynamics and Toyota Research Institute & 2024 & $\times$ \\
    \midrule
    
    \multirow{4}{=}{\centering Communication Agent} & 
    \multicolumn{2}{>{\centering\arraybackslash}m{0.34\textwidth}}{Semantic-driven AI Agent Communication} & UESTC & 2025.10 & $\times$ \\
    \cmidrule{2-6}
    & \multicolumn{2}{>{\centering\arraybackslash}m{0.34\textwidth}}{Agentic AI-enhanced SemCom} & BUPT & 2025.12 & $\times$ \\
    \cmidrule{2-6}
    & \multicolumn{2}{>{\centering\arraybackslash}m{0.34\textwidth}}{ChannelGPT} & BUPT & 2024 & $\times$ \\
    \cmidrule{2-6}
    & \multicolumn{2}{>{\centering\arraybackslash}m{0.34\textwidth}}{UniClaw} & China Unicom & 2026.3 & $\times$ \\
    \midrule

    \multirow{4}{=}{\centering Network Agent} & 
    \multicolumn{2}{>{\centering\arraybackslash}m{0.34\textwidth}}{RAN Agent} & Huawei & 2026.3 & $\times$ \\
    \cmidrule{2-6}
    & \multicolumn{2}{>{\centering\arraybackslash}m{0.34\textwidth}}{Agentic AI for RAN} & UPM & 2025.11 & $\times$ \\
    \cmidrule{2-6}
    & \multicolumn{2}{>{\centering\arraybackslash}m{0.34\textwidth}}{JoinAI-Agent} & China Mobile & 2026.2 & \checkmark \\
    \cmidrule{2-6}
    & \multicolumn{2}{>{\centering\arraybackslash}m{0.34\textwidth}}{Xingchen Super Agent} & China Telecom & 2025.9 & $\times$ \\
    \midrule

    \multirow{10}{=}{\centering Application Agent} & 
    \multirow{2}{=}{\centering General Agent} & Gemini 3 & Google & 2025.11 & $\times$ \\
    \cmidrule{3-6}
    & & Doubao-Seed-2.0 & ByteDance & 2026.02 & $\times$ \\
    \cmidrule{2-6}
    & \multirow{2}{=}{\centering Smart Factory} & Industrial Copilot & Siemens & 2024 & $\times$ \\
    \cmidrule{3-6}
    & & Joyindustrial & JD & 2025.5 & $\times$ \\
    \cmidrule{2-6}
    & \multirow{2}{=}{\centering Smart Healthcare} & Sully.ai & Sully.ai & 2025.6 & $\times$ \\
    \cmidrule{3-6}
    & & Hippocratic AI & Hippocratic AI & 2024 & \checkmark \\
    \cmidrule{2-6}
    & \multirow{2}{=}{\centering Smart City} & NVIDIA Blueprint for smart city AI & NVIDIA & 2025.6 & $\times$ \\
    \cmidrule{3-6}
    & & City Intelligent Agent Solution & iSSTech and Huawei & 2026.3 & $\times$ \\
    \cmidrule{2-6}
    & \multirow{2}{=}{\centering Intelligent Transportation} & TrafficGo & Huawei Cloud & 2019 & $\times$ \\
    \cmidrule{3-6}
    & & NaviAgent & Amap & 2025.6 & \checkmark \\
    \bottomrule
    \end{tabular}
\end{table}

\subsection{Embodied Agent}
\begin{itemize}
\item \textbf{SayCan:} Google proposes the SayCan\cite{brohan2023can}, an agent that integrates the high-level semantic planning capabilities of LLMs with the learned affordances of a robot's pre-trained skills. By grounding LLM outputs in feasible actions that align with both the robot's physical capabilities and the current environment, SayCan addresses key limitations of LAMs in real-world deployment: the lack of grounding, the generation of infeasible instructions for robots, and the difficulty of completing embodied long-horizon tasks described in abstract natural language. In real-world evaluations, SayCan demonstrates strong planning and execution success rates, nearly doubling the performance of non-grounded baseline models. It also supports zero-shot execution of long-horizon instructions, accommodates new skill additions, enables CoT reasoning and multilingual queries, and maintains interpretability throughout the decision-making process.

\item \textbf{Atlas:} Boston Dynamics, in collaboration with the Toyota Research Institute, has developed Large Behavior Models (LBMs) for the Atlas humanoid robot\footnote{https://bostondynamics.com/products/atlas/}. Trained on human demonstration data, the model employs a universal neural network and an end-to-end AI control strategy, eliminating the need for manual programming to adapt to specific scenarios. 
Atlas is capable of autonomously performing a variety of tasks, including running, jumping, obstacle crossing, and complex object manipulation. It can rapidly and intelligently recover from unexpected disturbances such as being pushed, and achieves movement speeds 1.5 to 2 times faster than humans, significantly enhancing the flexibility and adaptability of task execution.
\end{itemize}

\subsection{Communication Agent}
\begin{itemize}
\item \textbf{Semantic-driven AI Agent Communication:} Yu \textit{et al.} \cite{yu2025semantic} propose a semantic-driven AI agent communication framework comprising three stages: perception-aware semantic sampling, joint semantic-channel coding, and semantic resource orchestration. They develop three core enabling technologies: semantic adaptive transmission based on sample fine-tuning, semantic lightweight transmission integrating pruning, quantization, and partial sampling, and semantic self-evolution control employing a distributed multi-timescale hierarchical deep reinforcement learning method. Simulation results across three typical scenarios, including edge-to-edge, edge-to-BS, and multi-agent communication networks, demonstrate that the proposed scheme achieves faster convergence and greater robustness compared to traditional methods, with the distributed hierarchical optimization approach significantly outperforming conventional decision-making schemes.
\item \textbf{Agentic AI-enhanced SemCom:} Gao \textit{et al.} \cite{gao2025agentic} propose a unified agentic AI-enhanced SemCom framework consisting of an application layer, a semantic layer, and a cloud-edge collaboration layer. They also design the agentic KB-JSCC scheme, in which the source knowledge base is constructed by LLM and LVM agents, while the channel knowledge base is implemented by reinforcement learning agents. This solution addresses limitations of traditional bit transmission, which cannot adapt to new demands such as multi-agent collaboration in 6G scenarios, and overcomes defects in conventional SemCom, including limited representational capacity, index fragility at low SNR, insufficient multi-modal feature fusion, and lack of channel adaptivity. The proposed framework supports various 6G application scenarios and enables intelligent optimal scheduling of communication resources. 
\item \textbf{ChannelGPT:} Yu et al. propose ChannelGPT\cite{yu2025channelgpt}, a large model-driven digital twin channel generator embedded with environment intelligence. It adapts to 6G scenarios through a three-layer architecture and core capabilities such as multi-modal fusion and multi-task processing. Experimental validation demonstrates excellent performance in channel prediction accuracy and multi-scenario generalization, providing intelligent support for decision-making across all layers of 6G networks.
\item \textbf{UniClaw:} China Unicom has launched UniClaw\footnote{https://client.sina.com.cn/2026-03-06/doc-inhpzvnm9184324.shtml}, an AI-native communication capability centered on the core concept of ``AI reshapes connectivity" and based on China Unicom's Yuanjing digital intelligence capabilities.
UniClaw upgrades traditional basic communication functions such as telephony and short messaging to native-level AI connection channels. It aims to overcome the lack of AI empowerment in traditional basic communication, the high thresholds for intelligent service access, and insufficient multi-scenario responsiveness and security assurance. By enabling barrier-free access to intelligent services, stable multi-scenario response, and high-level security assurance, UniClaw positions the basic communication network as a key gateway in the era of intelligent agents.
\end{itemize}

\subsection{Network Agent}

\begin{itemize}
\item \textbf{RAN Agent:}
Huawei has released the industry's first RAN Agent\footnote{https://www.huawei.com/en/news/2026/3/mwc-mbb-ran}, built upon a specialized communication LAM and the Radio Digital Twin System (RDTS). This agent establishes an end-to-end efficient collaborative architecture and enables full-closed-loop intelligent operation, addressing critical limitations in traditional wireless networks including insufficient global intelligence, inaccurate network resource scheduling, and weak cross-scenario collaboration capabilities. The RAN Agent achieves precise network resource scheduling and full-scenario single-domain autonomy, delivering comprehensive improvements in user experience, operation and maintenance efficiency, and network energy efficiency while meeting diverse operator requirements across different deployment scenarios.
\item \textbf{Agentic AI for RAN:}
Pellejero \textit{et al.}~\cite{pellejero2025agentic} propose applying the Agentic AI paradigm to 5G/6G RAN management and optimization, integrating design patterns such as reflection and planning to enable autonomous decision-making through LAMs and multi-agent collaboration. This approach addresses the high complexity of next-generation networks, the inefficiency of traditional manual and static optimization methods, and the lack of mature frameworks in this domain. The proposed system achieves autonomous KPI monitoring, anomaly diagnosis, and optimization recommendations, with early industrial implementations demonstrating improved network performance and reliability.
\item \textbf{JoinAI-Agent:}
China Mobile has open-sourced the "JoinAI-Agent" intelligent agent engine\footnote{https://github.com/opencmit/JoinAI-Agent?tab=readme-ov-file}. Featuring a "one master with multiple slaves" architecture and no-code extension capabilities, it supports automation of complex enterprise processes. The engine has achieved top ranking in the international GAIA benchmark, breaking technical barriers, lowering development thresholds, and facilitating the construction of an open intelligent agent ecosystem while accelerating industrial intelligent transformation.
\item \textbf{Xingchen Super Agent:}
China Telecom has launched the Xingchen Super Agent\footnote{https://www.chinatelecom-h.com/en/cg/pdf/esg/innovation.pdf}, based on its self-developed Xingchen LAM. The agent incorporates core capabilities including autonomous task decomposition, cross-application collaboration, human-machine cooperation, and open customization, complemented by triple-layer security protection and a full-link evolution closed loop. This solution addresses the challenges of AI implementation in government and enterprise scenarios, improving business efficiency and decision-making quality while facilitating digital transformation and upgrading.
\end{itemize}

\subsection{Application Agent}

\subsubsection{General Agent}

\begin{itemize}
\item \textbf{Gemini 3:}
Google has unveiled Gemini 3\footnote{https://gemini3.us/gemini-3}, a new-generation flagship multi-modal intelligent agent model that significantly advances capabilities in deep reasoning, multi-modal integration, and long-term planning. This release addresses key limitations, including inaccurate understanding of user intentions, high barriers to entry for developers in intelligent agent development, and insufficient model robustness against adversarial inputs. 
\item \textbf{Doubao-Seed-2.0:}
Doubao-Seed-2.0\footnote{https://lf3-static.bytednsdoc.com/obj/eden-cn/lapzild-tss/ljhwZthlaukjlkulzlp/seed2/0214/Seed2.0\%20Model\%20Card.pdf} has undergone comprehensive optimization and upgrading to meet the practical demands of large-scale production deployments. Leveraging core capabilities in efficient reasoning, multi-modal understanding, and complex instruction processing, it offers three general-purpose Agent models (Pro, Lite, Mini) alongside a dedicated programming-oriented Code model.
\end{itemize}

\subsubsection{Vertical Agent}

\paragraph{Smart Factory}
\begin{itemize}
\item \textbf{Industrial Copilot:}
Siemens has developed Industrial Copilot\footnote{https://www.siemens.com/en-us/company/insights/generative-ai-industrial-copilot/}, an intelligent agent system built upon industrial infrastructure models and industrial agents, while simultaneously fostering an open ecosystem. Industrial Copilot leverages the integration of software and hardware, high-quality industrial data, and cross-domain industry knowledge to comprehensively address all facets of the industrial value chain. It supports engineers throughout end-to-end collaborative tasks, resolving persistent challenges such as the difficulty of implementing industrial AI and insufficient understanding of industrial logic. Currently serving over 200 customers, Industrial Copilot is projected to increase production efficiency by 50\%, achieve energy-saving optimization across multiple scenarios, and drive the transformation and upgrading of the manufacturing industry.
\item \textbf{Joyindustrial:}
JD's industrial LAM, Joyindustrial\footnote{https://jdcorporateblog.com/jd-industrials-unveils-joy-industrial-the-first-ai-model-designed-for-industrial-supply-chain-transformation/}, is optimized jointly for cost, efficiency, and user experience. By selecting the smallest-scale model based on ScalingLaw principles and employing a T-S strategy to reduce costs, adopting a Mixture-of-Experts (MoE) architecture and CoT training to improve efficiency, and optimizing user experience through domain data synthesis and reward function design, Joyindustrial achieves costs only one-sixteenth those of general-purpose LAMs while delivering an eightfold improvement in inference throughput. Furthermore, Joyindustrial enables the construction of agents capable of addressing diverse challenges within industrial ultra-long supply chains, including data silos, fragmented standards, complex management requirements, and collaborative conflicts.
\end{itemize}

\paragraph{Smart Healthcare}
\begin{itemize}
\item \textbf{Sully.ai:}
 Sully.ai\footnote{https://www.sully.ai/} is an enterprise-level AI assistant purpose-built for the healthcare industry, integrating advanced natural language processing technologies to comprehensively address core clinical and administrative workflows, including medical documentation, clinical research, administrative tasks, and multilingual translation. The platform seamlessly integrates with over ten electronic health record (EHR) systems, alleviating critical industry pain points such as the burden of manual documentation, inefficient information retrieval, cross-language communication barriers, and low institutional operational efficiency. Sully.ai enhances both the quality of medical services and institutional operational efficiency while eliminating language barriers in doctor-patient communication. Recognized by over 400 medical institutions, it provides robust support for the digital transformation of the healthcare industry.
\item \textbf{Hippocratic AI:}
Hippocratic AI\footnote{https://www.trially.ai/} has launched an AI-powered medical workforce platform built upon its proprietary Polaris model. Designed with safety, multilingual accuracy, and seamless electronic medical record system integration as foundational priorities, the platform delivers multilingual non-diagnostic routine nursing services. It addresses critical global challenges including the shortage of nursing resources, the burden of repetitive tasks on medical staff, the high costs of traditional care delivery, and the inability of general-purpose AI systems to meet stringent medical compliance requirements. Hippocratic AI achieves high clinical accuracy and patient satisfaction, with deployments across multiple medical institutions handling massive volumes of medical calls while reducing costs and improving efficiency. The platform has secured substantial financing and gained significant market recognition.
\end{itemize}

\paragraph{Smart City}
\begin{itemize}
\item \textbf{NVIDIA Blueprint for Smart City AI:}
NVIDIA has introduced a smart city AI blueprint\footnote{https://www.nvidia.com/en-us/industries/smart-cities-and-spaces/} built on OpenUSD digital twins and tools such as Omniverse. By deploying AI agents through a three-stage workflow, this blueprint addresses the challenges of infrastructure lag and inefficient urban operations resulting from rapid population growth in cities. Based on this blueprint, cities can deploy an integrated operational platform that combines weather data, traffic sensors, and emergency response systems, helping optimize response speed, infrastructure planning, real-time monitoring, and other urban capabilities, thereby facilitating the transformation of urban operations.
\item \textbf{City Intelligent Agent Solution:}
iSSTech and Huawei have jointly developed a city intelligent agent solution\footnote{https://e.huawei.com/jp/news/2026/industries/government/city-intelligent-agent} that integrates AI models and big data technologies within a five-layer architecture centered on converged intelligence. The solution covers three major application scenarios: city governance, economic development, and public services. It addresses critical bottlenecks in urban intelligence development, including weak infrastructure and insufficient localized AI capabilities. The solution reduces project cycles by 50\%, streamlines government affairs, and helps cities advance their multi-domain intelligence capabilities while building sustainable localized AI capacity.
\end{itemize}

\paragraph{Intelligent Transportation}
\begin{itemize}
\item \textbf{TrafficGo:}
Huawei Cloud has launched the TrafficGo intelligent transportation solution\footnote{https://www.huaweicloud.com/product/trafficgo.html}, which integrates multi-source data with AI, edge computing, and other technologies to build an intelligent traffic system. It enables functions such as regional signal coordination and intelligent congestion management, ensuring smooth traffic flow and improving overall efficiency.

\item \textbf{NaviAgent:}
Amap introduces NaviAgent\footnote{https://www.alibabagroup.com/en-US/document-1889126073686294528}, the world’s first AI navigation agent in the map field. It adopts a planner-executor architecture and a smart closed-loop system comprising four modules, integrating traffic perception, emotion-aware voice interaction, and other technologies. NaviAgent addresses the limitations of traditional navigation, such as restricted local perception, rigid execution, and lack of emotional interaction, by enabling beyond-visual-range road condition prediction, lane-level safety warnings, and emotional companionship. This transforms navigation from a mere tool into an intelligent travel partner.
\end{itemize}

\section{Challenges and Future Research Directions}
\label{sectionfuture}

This section delves into the challenges and future research directions of semantic-based agent communication networks as follows.

\subsection{Theoretical Framework of Semantic-based Agent Communication Networks}

Classical Shannon information theory measures information in bits, but it cannot characterize the ``meaning" and ``value" at the semantic level. There is a lack of a unified mathematical foundation for defining a measurement unit for semantic information to quantify its impact on the decision-making of the receiving agent. Furthermore, SemCom improves bandwidth efficiency through the integration of communication, computation, and intelligence, yet existing theories cannot characterize the performance boundaries of this joint optimization. Moreover, noise and interference in wireless channels may lead to semantic misunderstandings. Establishing a mathematical model for semantic channels to describe the loss or deviation of semantic information during transmission is also important. Future research can integrate the fundamental theories of AI to further enrich the theory of semantic-based agent communication networks. In addition, theoretical support can be provided from an interdisciplinary perspective, such as theories from complex networks and systems science.

\subsection{Management of Semantic KBs for Semantic-based Agent Communication Networks}
Different agents may construct semantic KBs based on different foundation models, KGs, or domain knowledge. Therefore, when they interact for the first time, achieving fast and accurate alignment of semantic KBs to ensure the matching of encoding and decoding poses a significant challenge. Furthermore, as world knowledge continuously evolves, agents' semantic KBs must undergo ongoing evolution, raising the following issues: How to efficiently update semantic KBs? How to prevent old semantic knowledge from interfering with new semantic knowledge? How to design efficient and highly reliable forgetting mechanisms for semantic knowledge? Additionally, on agents with constrained computational power and storage, the storage and query overhead of semantic KBs must be minimized. However, existing large-scale KGs or neural network parameters are difficult to deploy directly, necessitating extreme lightweight semantic representations. Therefore, future research can focus on the following directions: lightweight semantic knowledge representation, fast alignment of semantic KBs, and dynamic evolution of semantic KBs.

\subsection{Security and Privacy Protection for Semantic-based Agent Communication Networks}

Compared to traditional bit-level communication, SemCom may introduce new security threats to agent communications. For example, attackers no longer need to corrupt all data. On the contrary, they only need to tamper with a small amount of critical semantics to completely alter the meaning of the information and mislead the agent's decision-making. Furthermore, if an agent generates erroneous semantic information due to model hallucination and it spreads rapidly through the SemCom network, it can lead to a systemic false consensus, undermining collaborative trust. To secure semantic-based agent communication networks, future directions include: researching the representation of encrypted semantics while maintaining semantic usability between legitimate agents; introducing technologies such as blockchain and digital signatures to add traceable and tamper-proof evidence to semantic information, enabling the receiving agent to verify the authenticity and integrity of semantics; utilizing trusted execution environments and secure multi-party computation to protect data and model privacy during semantic encoding and decoding processes, ensuring that semantic information is not leaked even if computing nodes are untrusted; and designing interpretable semantic models to enhance the detection capability of potential attacks.

\subsection{Standardization and Industry for Semantic-based Agent Communication Networks}
The standardization of SemCom has been advanced in numerous standardization organizations, such as the ITU, 3GPP, International Mobile Telecommunications 2030 (IMT-2030), and China Communications Standards Association (CCSA). Semantic-based agent communication, as one of its application scenarios, has garnered attention from both academia and industry. For example,  Future directions include: further promoting semantic-based agent communication from the perspective of the entire industry chain, encompassing fundamental theories, key technologies, prototype development, chip manufacturing, and application deployment; defining cross-domain and cross-industry semantic translation standards to achieve semantic interoperability across vertical industries; establishing open-source platforms and testing environments for the validation of new algorithms and models; and developing agent communication network technologies that enable the coexistence of semantics and bits, adapting to network environments where semantics and bits are mixed.

\section{Conclusions}
\label{sectionconclusion}
In this review, we have presented a comprehensive framework for semantic-based agent communication networks, addressing the critical intersection of SemCom and agentic AI systems. We first proposed a novel architecture for semantic-based agent communication networks, comprising three layers, four entities, and four stages. It establishes a structured foundation for understanding and designing semantic-enabled agent communication networks. This architecture integrates three wireless agent network layers with four AI agent entities and four operational stages that form a complete cognitive cycle for agent behavior. Building upon this architectural framework, we conducted an extensive exploration of the state-of-the-art in semantic-based agent communication networks. Our investigation spanned the three architectural layers, examining representative approaches from intention inference techniques to semantic coding and distributed collaboration mechanisms. Furthermore, we systematically reviewed advancements across the four stages, including perception, memory, reasoning, and action, highlighting how semantics enhance each stage. We also provided a taxonomy of AI agents in SemCom networks, categorizing them into embodied, communication, network, and application agents to clarify their distinct roles and functionalities. Despite significant progress in this emerging field, several fundamental challenges remain open for future investigation. Lastly, we identified and discussed key research directions.

%
%
%
%

	\Acknowledgements{This work was supported in part by the National Key Research and Development Program of China (Grant No. 2020YFB1806905); in part by the National Natural Science Foundation of China (Grant Nos. 62501066 and U24B20131); and in part by the Beijing Municipal Natural Science Foundation (Grant No. L242012).}

\bibliography{references}
\bibliographystyle{unsrt}

	
%
%
%


        



	
\end{document}